\documentclass[twocolumn,showpacs,amsmath,amssymb,prc,superscriptaddress,nofootinbib]{revtex4-1}

\usepackage{graphicx}

\usepackage{dcolumn}
\usepackage{bm}
\usepackage{times}
\usepackage{siunitx}
\usepackage{booktabs}
\usepackage{etoolbox}
\usepackage{braket}
\usepackage{threeparttable}
\usepackage{caption}
\usepackage{subcaption}
\usepackage{multirow}
\usepackage{xcolor}
\robustify\itshape

\AtBeginDocument{
\heavyrulewidth=.08em
\lightrulewidth=.05em
\cmidrulewidth=.03em
\belowrulesep=.65ex
\belowbottomsep=0pt
\aboverulesep=.4ex
\abovetopsep=0pt
\cmidrulesep=\doublerulesep
\cmidrulekern=.5em
\defaultaddspace=.5em
}

\newcolumntype{.}{D{.}{.}{1}}
\newcolumntype{X}{D{X}{X}{1}}

\newcommand{\printthis}[2][true]{%
\ifbool{#1}{%
#2}{}}

\begin{document}
\def\sun{\odot}

\title{Bayesian Analysis of the $^{70}$Zn$(d, ^3 \textnormal{He}) ^{69}$Cu Transfer Reaction}
 
\author{C.~Marshall} \affiliation{Department of Physics, North
  Carolina State University, Raleigh, NC 27695, USA}
\affiliation{Triangle Universities Nuclear Laboratory, Durham, NC
  27708, USA}

\author{P.~Morfouace} \affiliation{CEA, DAM, DIF, F-91297 Arpajon, France}

\author{N. de S\'{e}r\'{e}ville} \affiliation{Institut de Physique Nucl ́eaire et Universit ́e Paris-Sud, 91406 Orsay Cedex, France}

\author{R.~Longland} \affiliation{Department of Physics, North
  Carolina State University, Raleigh, NC 27695, USA}
\affiliation{Triangle Universities Nuclear Laboratory, Durham, NC
  27708, USA}

\begin{abstract}
  Transfer reactions provide information about the single-particle nature of nuclear levels. In particular,
  the differential cross sections from these measurements are sensitive to the angular momentum of the transferred particle
  and the spectroscopic factor of the populated level. However, the process of extracting these properties is
  subject to uncertainties, both from experimental and theoretical sources. By integrating the distorted wave Born approximation into a Bayesian model, we propagate
  these uncertainties through to the spectroscopic factors and orbital angular momentum values. We use previously reported data of the proton pickup reaction $^{70}$Zn$(d, ^3 \textnormal{He}) ^{69}$Cu as an example.   
  By accounting for uncertainties in the experimental data, optical model parameters, and reaction mechanism, we
  find that the extracted spectroscopic factors for low lying states of $^{69}$Cu are subject to large, asymmetric uncertainties ranging from $35 \%$ to $108 \%$. Additionally,
  Bayesian model comparison is employed to assign probabilities to each of the allowed angular momentum transfers. This method confirms the assignments for many
  states, but suggests that the data for a state lying at $3.70$ MeV is better characterized by an $\ell = 3$ transfer, rather than the previously reported $\ell = 2$.
\end{abstract}


\maketitle


\section{Introduction}

Nucleon transfer reactions are critical tools for studying the single particle structure in nuclei.
However, their usefulness depends upon a successful theoretical description of the reaction mechanism.
By far the most widely used methodology is the combination of the nuclear    
optical model \cite{feshbach1958} and the \underline{D}istorted \underline{W}ave
\underline{B}orn \underline{A}pproximation (DWBA) (see Ref.~\cite{sum_rules} and references therein).
By using DWBA it becomes possible to determine both the transferred angular momentum, $\ell$, and
spectroscopic factor of the populated single particle or hole state \cite{satchler}.  
This structure information can in turn be used to answer questions in nuclear astrophysics \cite{transfers_in_astro},
and to test the shell model on isotopes located far from stability \cite{Wimmer_2018}. 

Despite the wide use of these methods, quantifying the uncertainties associated with
both the optical potentials and the reaction model has been a long standing issue. Previous studies have used statistical methods to
determine the uncertainty on the potential parameters \cite{varner}, but little work has been done to propagate
these uncertainties through DWBA calculations in a statistically meaningful way. To data, most spectroscopic factors are reported
with either no uncertainty, an assumed equivalence between the uncertainty in the data normalization and that of the spectroscopic factor, or
a constant $25 \%$ determined from historical studies \cite{endt_cs}.

Over the last few years, these issues have led to a renewed focus on the impact of optical model parameters on transfer reactions. A series of studies has focused on the nature and magnitude of this effect \cite{Lovell_2015, lovell_opt_unc, king_d_p}. The first steps have also been taken towards quantifying these uncertainties using Bayesian statistics \cite{lovell_mcmc, king_dwba}. These studies focus on the broad effects of optical potentials, but it is worthwhile to establish a
Bayesian framework in which the results of a single experiment can be analyzed. The goal of this paper is to establish such a framework and to examine
the possible implications on future experiments.

The methods developed and presented here will be applied to the analysis of
the proton pickup reaction $^{70} \textnormal{Zn} (d, ^{3} \textnormal{He}) ^{69} \textnormal{Cu}$,
which was originally reported in Ref. \cite{pierre_paper}.
This data set possesses many of the features typical of a transfer measurement study: the use of a high resolution magnetic spectrograph
to resolve the excited states of interest, elastic data for the entrance channel
collected with the same target and beam, experimental uncertainties coming from counting statistics,
and limited angular coverage in both the elastic scattering and transfer
differential cross sections. The previous analysis assigned $\ell$ values and extracted spectroscopic factors for
the first eight excited states of $^{69} \textnormal{Cu}$. Our reanalysis aims to determine the uncertainties associated
with these quantities using Bayesian statistics.

This paper will be structured to introduce the relevant reaction theory in Sec.~\ref{sec:reaction_theory}, 
explain and construct the Bayesian model in Sec.~\ref{sec:bayes_stuff}, and finally present and
then discuss the results in Sec.~\ref{sec:results} and Sec.~\ref{sec:discussion}, respectively.

\section{Reaction Theory}
\label{sec:reaction_theory}

\subsection{The Optical Model}
\label{sec:op}

The nuclear optical model simplifies the multi-nucleon scattering problem by
considering a single particle interacting with a complex potential, $\mathcal{U}(r)$. The theoretical basis for this procedure was first
established in Ref.~\cite{feshbach1958}, but fell short of actually prescribing
the form of the complex potential. Through detailed analysis of elastic scattering
from a range of targets and energies, Ref.~\cite{b_and_g_d} developed a phenomenological form for the optical model. Our work will be focused on the
effect of these phenomenological potential parameters, but further theoretical and historical details can be found in Ref.~\cite{hodgson1971}.
For our work we adopt the following form of the optical potential:

\begin{multline}
  \label{eq:optical_model}
  \mathcal{U}(r) = V_c(r; r_c)-Vf(r; r_0, a_0) \\
  -i(W-4a_iW_s\frac{d}{dr_i})f(r; r_i, a_i) \\
  + (\frac{\hbar}{m_{\pi}c})^2V_{so} \frac{1}{r} \frac{d}{dr} f(r; r_{so}, a_{so}) \boldsymbol{\sigma}  \cdot \boldsymbol{\ell}, 
\end{multline}

where $f(r)$ is given by the Wood-Saxon form factor:

\begin{equation}
  \label{eq:ws_pot}
  f(r; r_0, a_0) = \frac{1}{1 + \exp(\frac{r-r_0A_t^{1/3}}{a_0})}.
\end{equation}

Each term in $\mathcal{U}$ is parameterized with a well depth, ${V, W, W_s}$, radius, ${r_0, r_i, r_{so}}$, and diffuseness, ${a_0, a_i, a_{so}}$. $A_t$ is the mass number of
the target nucleus.
Additionally, the spin orbit term has an interaction based on the projectile orbital and spin angular momentum, $\boldsymbol{\ell}$ and $\bf{s}$, respectively. In
this case $\boldsymbol{\sigma} = 2 \bf{s}$, and $(\frac{\hbar}{m_{\pi}c})^2$ is a constant with a value of approximately $2$ fm$^2$.
The Coulomb term, $V_c$, comes from the potential of a uniformly charged sphere with radius $R_c = r_c A_t^{1/3}$. 
These conventions are adopted in order to be consistent with the code FRESCO \cite{fresco}, and care should be taken to convert values
given in this paper if a different set of conventions is adopted.

The phenomenological optical model uses experimental data, typically differential elastic scattering cross sections and analyzing powers, to determine the parameter values defined above. Local fits of these parameters, like those listed in Ref.~\cite{perey_perey}, try to best reproduce the results of elastic scattering from a single target nucleus at a single energy. Global fits, such as Refs.~\cite{varner, b_and_g_d, b_g_p, b_g_3he, pang_global, daehnick_global}, use a variety of targets and beam energies to derive relations between potential parameters and target mass, beam energy, and other nuclear properties.         

\subsection{Distorted Wave Born Approximation}
\label{sec:DWBA}

The shape and magnitude of transfer reaction differential cross sections are directly related to
the angular momentum of the transferred nucleon and the spectroscopic factor of the populated nuclear state.
Thus, a theoretical description of the cross section will
allow the extraction of these properties from experimental data. DWBA is a perturbative method that uses the optical potentials of
Sec.\ref{sec:op} to model the entrance and exit channels, and a {transition operator} for the transferred particle
or cluster. Expressing this {transition operator} explicitly for the $A(d, ^3 \! \textnormal{He})B$ pickup reaction, we can write
it in either the prior or post form:

\begin{align}
  \label{eq:eff_pot}
  & \mathcal{V}_{\textnormal{prior}} = V_{p+d} + \mathcal{U}_{d+B} - \mathcal{U}_{d + A} \\
  & \mathcal{V}_{\textnormal{post}} = V_{p+B} + \mathcal{U}_{d+B} - \mathcal{U}_{^3\textnormal{He} + B}.                                                                             
\end{align}

$\mathcal{U}$ are the optical potentials for each of the reaction channels.
The entrance, exit, and core-core systems are denoted by $A + d$, $B + ^{3} \textnormal{He}$, and $B + d$, respectively.
The $V$ potentials are the binding potentials for the
proton on either the projectile or target nucleus. The first order $T$ matrix for the
transfer from channel $\alpha$ to channel $\beta$ in the prior form is given by:

\begin{multline}
  \label{eq:2}
  T_{\beta \alpha} = J \int d \boldsymbol{r}_{^3\textnormal{He}}  \int d \boldsymbol{r}_d \\
  \chi_{\beta}^{(-)*}(\boldsymbol{r}_{^3\textnormal{He}}, \boldsymbol{k}_{^3\textnormal{He}})  \bra{B, ^3\textnormal{He}} \mathcal{V}_{\textnormal{prior}}  \ket{A, d} \chi_{\alpha}^{(+)}(\boldsymbol{r}_{d}, \boldsymbol{k}_{d}), 
\end{multline}

where $\chi$ is the distorted wave generated from the corresponding optical potential, {J is the Jacobian for the transformation to the two coordinates}, and the kets refer to the internal coordinates
of the respective nuclei. Further information on the derivation of these equations and other theoretical considerations
can be found in Refs. \cite{satchler, thompson_nunes_2009}.

\subsection{Cross Section Calculations}
\label{sec:cs_calc}

All transfer and elastic differential cross sections for this work were calculated using
the coupled-channels reaction code FRESCO \cite{fresco}.
The FRESCO transfer differential cross section can be related to experiment through:

\begin{equation}
  \label{eq:spec_factor}
  \frac{d \sigma}{d \Omega}_{exp} = C^2S_{\textnormal{p}} C^2S_{\textnormal{t}} \frac{d \sigma}{d \Omega}_{FRESCO}.
\end{equation}

The two spectroscopic factors and isospin Clebsch-Gordan coefficients are for the projectile and target system, respectively. For
particles with $A \leq 4$, the spectroscopic factor can be approximated by $\frac{A}{2}$ \cite{satchler}. Thus, for the $d + p$ system, $C^2S_{\textnormal{p}} = \frac{3}{2}$. Since $C^2S_{\textnormal{p}}$ is assumed constant, any further reference to $C^2S$ will be equivalent to $C^2S_{\textnormal{t}}$.  

In order to reduce the computational cost of the transfer calculations,
this work uses the zero-range approximation  \cite{satchler}. This approximation, in the specific case of the pick-up reaction $A(d, ^{3} \! \textnormal{He})B$, takes the prior form of Eq.~(\ref{eq:eff_pot}) and sets $\mathcal{U}_{d + B} - \mathcal{U}_{d+A}$ to zero, a procedure justified by experimental observation \cite{first_dwba}. The projectile is then assumed to be absorbed and emitted from the same point giving:

\begin{equation}
  \label{eq:1}
   \bra{d} V_{pd} \ket{^3\textnormal{He}} \sim D_o \delta(\mathbf{r_p}) , 
\end{equation}

where $\ket{^3\textnormal{He}}$, $\ket{d}$ are the internal wave functions of the ejectile and projectile, respectively, $D_0$ is
the volume integral of the interaction strength, $V_{pd}$ is the binding potential of the proton to the deuteron,
and $\mathbf{r_p}$ is the coordinate of the proton relative to the deuteron. Use of this approximation gives us the further benefit of a direct
comparison to the original analysis of $^{70}$Zn$(d, ^3 \textnormal{He}) ^{69}$Cu that used the zero-range code DWUCK4 for the extraction of $C^2S_t$ \cite{dwuck4}. It should be noted that Ref.\cite{pierre_paper} also performed finite-range calculations, but the computational costs are prohibitively expensive in the present analysis. {The value of $D_0$ is calculated
  theoretically, with the historical value for proton pick-up and stripping reactions being $D_0 = -172.8$ MeV fm$^{3/2}$ \cite{bassel}}. Comparing the different models in Ref.~\cite{all_norms_3He}, an approximately $15 \%$ spread in the values of $D_0^2$ is observed. This is inline with the findings of Ref.~\cite{bertone}, which also noted an approximate $15 \%$ spread in the product $(C^2S_{\textnormal{p}}) D_0^2$. {We adopt the above value with its associated uncertainty; however, \textit{Ab initio} methods, such as those in Ref.~\cite{brida_ab_initio}, now offer more precise determinations of the $\braket{d|^3 \textnormal{He}}$ overlap. If $D_0$ is deduced using these methods, then this additional source of uncertainty will be effectively eliminated. As a direct consequence the uncertainty in $C^2S_{\textnormal{t}}$ will be reduced by about $15 \%$.}

\section{Bayesian Inference}
\label{sec:bayes_stuff}

As discussed in Sec.~\ref{sec:op}, the potential parameters of the phenomenological optical model
are constrained by fitting to experimental data. Thus, their values
are inherently subject to uncertainty, which will ultimately propagate through
to any quantity extracted using them. Bayesian inference treats this interaction
between measured data and model parameters as a logical relationship between
conditional probabilities \cite{bayes}. This relationship is expressed in  
Bayes' theorem:

\begin{equation}
  \label{eq:b_theorem}
  P(\boldsymbol{\theta}|\mathbf{D}) = \frac{P(\mathbf{D}|\boldsymbol{\theta}) P(\boldsymbol{\theta})}
  {\int_{\boldsymbol{\theta}} P(\boldsymbol{D}|\boldsymbol{\theta})P(\boldsymbol{\theta}) d\boldsymbol{\theta}},
\end{equation}

where the posterior probability distribution, $P(\boldsymbol{\theta}|\mathbf{D})$, is the conditional probability of the model parameters, $\boldsymbol{\theta}$, given the data, $\mathbf{D}$. The posterior is calculated from the likelihood function, $P(\mathbf{D}|\boldsymbol{\theta})$, prior probabilities, $P(\boldsymbol{\theta})$, and the evidence integral, $\int_{\boldsymbol{\theta}} P(\boldsymbol{D}|\boldsymbol{\theta})P(\boldsymbol{\theta}) d\boldsymbol{\theta}$. Prior probabilities represent our knowledge of the
parameters before the data is considered, and must be assigned for every parameter that we want to estimate. The likelihood function
must also be specified, and will express, in probabilistic terms, how the parameters of the model relate to the data. This function is also present in the frequentist approach, with a common example being the $\chi^2 $ function. The evidence ensures
that the product of the likelihood and the priors is normalized. For this work one of our main goals will be to estimate the posterior distribution
for the spectroscopic factor. This will require that we assign priors for every optical model potential parameter and the spectroscopic factor itself. These prior probabilities will then be updated through the likelihood function using the experimentally measured cross sections for the elastic and transfer channels.   

Bayes' theorem is also central to our other goal of determining the most probable angular momentum
transfer for a given state. This problem belongs to a subcategory of Bayesian inference called model selection. Computing the probability for a
model, $M_j$, can be done by restating Bayes' theorem:

\begin{equation}
  \label{eq:m_theorem}
  P(M_j|\mathbf{D}) = \frac{P(\mathbf{D}|M_j) P(M_j)}
  {\sum_i P(\mathbf{D}|M_i)P(M_i)}.
\end{equation}

\noindent This expression is built on the same logical foundation as Eq.~(\ref{eq:b_theorem}), but has been adapted to compute
posterior distributions for $M_j$, which means a comparison can now be made between different models. For each $M_j$ there is a set of model parameters $\boldsymbol{\theta}_j$ which have been marginalized over. This means:

\begin{equation}
  \label{eq:marg}
  P(\mathbf{D}|M_j) = \int P(\mathbf{D}|M_j, \boldsymbol{\theta}_j) P(\boldsymbol{\theta}_j|M_j) d\boldsymbol{\theta}_j.
\end{equation}

\noindent Based on this equation it can be seen that $P(\mathbf{D}|M_j)$ is equivalent to  
the evidence integral from Eq.~(\ref{eq:b_theorem}). Thus, in order for us to evaluate how probable different angular
moment transfers are, we must calculate the evidence integral.

Once the evidence integral is calculated, there are several metrics to interpret model posterior probabilities. For simplicity, we will now refer to the evidence integral as $Z_j$, which corresponds to the model $M_j$.
The most commonly used criterion for Bayesian model selection is called the Bayes Factor, which is defined by:

\begin{equation}
  \label{eq:bayes_factor}
  B_{ji} = \frac{Z_j}{Z_i}.
\end{equation}

\noindent If this ratio is greater than $1$, the data support the selection of model $j$, while values less than $1$
support model $i$. Judging the level of significance for a value of  $B_{ji}$ is open to interpretation, but a useful
heuristic was given by Jefferys \cite{Jeffreys61}. For the cases where model $j$ is favored over $i$ we have the following
levels of evidence: $3 > B_{ji} > 1$ is anecdotal, $10 > B_{ji} > 3$ is substantial, $30 > B_{ji} > 10$ is strong, $100 > B_{ji} > 30$ is very strong,
and $ B_{ji} > 100$ is decisive.

It is also possible to calculate explicit probabilities for each model. Assuming each of the models is equally likely, the probability of a given model can be expressed as:

\begin{equation}
  \label{eq:model_prob}
  P(M_j|\mathbf{D}) = \frac{Z_j}{\sum_i Z_i}.
\end{equation}

Through Eq.~(\ref{eq:model_prob}), probabilities can be calculated for each physically allowed angular momentum transfer, $\ell_j$.
Using these definitions Bayesian inference can be carried out after prior probabilities are assigned for each optical model parameter and
a likelihood function for the data is chosen.

\subsection{Ambiguities in Potential Parameters}
\label{sec:amb_pots}

Any analysis involving potentials of the form in Eq.~(\ref{eq:ws_pot}) will suffer from so-called continuous and discrete ambiguities.
Both of these ambiguities arise because a single differential cross section at a single energy cannot uniquely determine the potential parameters.
The continuous ambiguity describes strong correlation between certain model parameters \cite{hodgson1971, vernotte_optical}. A well known example is the relation between the real volume depth, $V$, and the corresponding radius, $r_0$.
The relation has an approximate analytical form given by $Vr_0^n = const$, where the exponent $n$ and the constant vary depending on the reaction. 
This issue can be remedied in part by a global analysis of the potential parameters
across a wide range of mass numbers and reaction energies, as noted in the comprehensive analysis of
proton and neutron scattering in Ref.~\cite{varner} and for $^3$He and $t$ scattering in Ref.~\cite{pang_global}.
Since our analysis will be limited to a single elastic scattering data set, our model must be prepared to deal
with these parameter correlations. We explicitly demonstrate the existence of these ambiguities for $^{70} \textnormal{Zn}(d, d)^{70} \textnormal{Zn}$ 
in Appendix A.

The discrete ambiguity arises in optical model analysis due to the identical
phase shifts that are produced by different values of $V$ \cite{drisko_1963}.
This multi-modal behavior is perhaps the more problematic of the two ambiguities
since parameter correlation can be handled with standard statistical methods. In particular, interpretation
of uncertainties in a multi-modal problem requires care beyond standard credibility intervals.
The discrete families of parameters can be readily identified by the volume integral of the real potential:

\begin{equation}
  \label{eq:j_int}
  J = \frac{4 \pi}{A_{P}A_{T}} \int_0^{\infty} Vf(r; r_0, a_0)  r^2 dr ,
\end{equation}

where the mass numbers of the projectile and target, $A_P$ and $A_T$, respectively, ensure
that $J$ should be roughly constant for a family of potential parameters at a single energy.
Microscopic structure models such as the folding model can also be used to calculate $J$,
and this theoretical value can be used to identify the physical potential family \cite{daehnick_global}.
Trusting the efficacy of this method, our approach for this work is to adopt potential
depths from global fits and to keep our prior values contained around these starting potential depths.

\subsection{Global Potential Selection}

The {initial} potentials used for the analysis of $^{70}$Zn$(d, ^3 \textnormal{He}) ^{69}$Cu before inference can be found in Table~\ref{tab:opt_parms}.
In order to facilitate comparison with Ref.~\cite{pierre_paper}, we have used the same global
potentials. In particular, we take the values of the Daehnick-F global $d$ optical model \cite{daehnick_global},
and the Becceheti and Greenless global $^3$He model of Ref.~\cite{b_g_3he}. It is also worth noting that elastic scattering with an
unpolarized beam does not provide a constraint on the parameters of a spin-orbit potential, so all spin orbit terms have been held fixed
in the current work \cite{hodgson1994, daehnick_global, thompson_nunes_2009}.

The bound state geometric parameters are assigned their most commonly used value of $r_0=1.25$ fm and $a_0=0.65$ fm, 
with the volume potential depth adjusted to reproduce the binding energy of the final state \cite{perey_params,hodgson1971,bjork_params}.
The bound state spin-orbit volume depth was fixed at a value of $V_{so} = 8.66$ MeV in order to
approximately correspond to the condition  $\lambda = 25$, where $\lambda \sim  \frac{180 V_{so}}{V}$
for the value of $V$ for the ground state. 

\begin{table*}[ht]
\begin{threeparttable}[e]
\caption{\label{tab:opt_parms}Optical potential parameters used in this work.}
\centering
\setlength{\tabcolsep}{4pt} 
\begin{tabular}{ccccccccccccccc}
\toprule[1.0pt]\addlinespace[0.3mm] Interaction  & $V$ & $r_{0}$ & $a_{0}$ & $W$ & $W_{s}$ & $r_{i}$ & $a_{i}$ & $r_{c}$ & $V_{so}$ \\
                                                 & (MeV) & (fm) & (fm) & (MeV) & (MeV) & (fm) & (fm) & (fm) & (MeV)\\ \hline\hline\addlinespace[0.6mm]
$d$ $+$ $^{70}$Zn\tnote{a} & $86.76$ & $1.17$ & $0.75$ & $0.90$  & $11.93$ & $1.32$  &  $0.81$ & $1.30$ & $6.34$ &  \\
\hspace{0.15cm} $^{3}$He $+ ^{69}$Cu \tnote{b} & $156.5$ & $1.20$ & $0.72$ & $42.2$ & &$1.40$ & $0.86$ & $1.25$ & \\

\hspace{0.1cm}$p$ $+$ $^{69}$Cu & \tnote{c} & 1.25 & 0.65 & & & & & 1.25 & 8.66 & \\[0.2ex]
\bottomrule[1.0pt]
\end{tabular}
\begin{tablenotes}
\item[a] Global potential of Ref. \cite{perey_perey_d}.
\item[b] Global potential of Ref. \cite{b_g_3he}.
\item[c] Adjusted to reproduce binding energy of the final state.
\end{tablenotes}
\end{threeparttable}
\end{table*}

\subsection{Bayesian Model}
\label{sec:model}

Following the above discussion and considerations, we will now define our Bayesian model, which fits each excited state simultaneously with the elastic scattering data. In order to do this, each
parameter, whether from the optical model potentials or otherwise, has to be assigned a prior probability distribution.
Additionally, likelihood functions will need to be assigned for the data in both the elastic and transfer channels.
For this work we will only need three distributions: normal, half-normal, and uniform. The normal distribution is
defined according to its location parameter, $\mu$, and scale parameter, $\sigma^2$. Symbolically this is given by
$\mathcal{N}(\mu, \sigma^2)$. A half-normal distribution is equivalent to a normal distribution with $\mu=0$ and restricted to the interval $[0, \infty)$.
We write it as $\textnormal{HalfNorm}(\sigma^2)$. Finally, the uniform distribution will be given by its lower limit and its upper limit, written
as $\textnormal{Uniform}(\textnormal{Lower}, \textnormal{Upper})$.

The majority of parameters come from the optical model potentials. The elastic scattering data from $^{70} \textnormal{Zn} (d,d)$ should be able
to inform the posteriors for the entrance channel parameters, $\boldsymbol{\mathcal{U}}_{\textnormal{Entrance}}$. However, the ambiguities discussed in Sec.~\ref{sec:amb_pots} 
combined with the lack of data at angles higher than $\theta_{c.\!m.}= 50^{\circ}$ means that the priors for the entrance channel must be weakly informative. In order to accomplish this, we focus the radius and diffuseness parameters for both the real and imaginary potentials around a reasonable range. If we assume that physical values for these parameters tend to lie within $r = 1.0-1.5$ fm and $a=0.52-0.78$ fm, then we can construct our priors to favor these values. This is accomplished by assigning normal distributions with locations, $\mu_r = 1.25$ fm and $\mu_a=.65$ fm and scale parameters $\sigma^2_r = (0.20 \, \mu_r)^2$ and $\sigma^2_a = (0.20 \, \mu_a)^2$. These priors have $68 \% $ credibility intervals that are equivalent to $r = 1.0-1.5$ fm and $a=0.52-0.78$ fm, and importantly do not exclude values that lie outside of these ranges. This means that if the data are sufficiently informative, they can pull the values away from these ranges, but in the absence of strong evidence our priors will bias the parameters toward their expected physical values. The depths of the potentials were also assigned scale parameters of $20 \% $ of their global depths. This favors the mode assigned by the global analysis, thereby eliminating the discrete ambiguity and producing a unimodal posterior. These conditions are summarized in the prior:

\begin{equation}
  \label{eq:entrance_prior}
  \boldsymbol{\mathcal{U}}_{\textnormal{Entrance}} \sim \mathcal{N}(\mu_{\textnormal{central}, k}, \{0.20 \, \mu_{\textnormal{central}, k}\}^2), 
\end{equation}

where the symbol $\sim$ denotes ``distributed according to", ``central" refers to the global values for the depths and the central physical values of $r=1.25$ fm and $a=0.65$ fm defined above, and the index $k$ runs over the depth, radius and diffuseness parameters for the real and imaginary parts of the potential.

The exit channel, as opposed to the entrance channel, does not have elastic scattering data to constrain it directly. This means that informative priors based on a global analysis must be used, while also considering a reasonable range of values. Normal priors are used, again to avoid sharp boundaries on the values, with the global values of Table~\ref{tab:opt_parms} as the location parameters, and the scale parameter set to $\sigma^2 = (0.10 \, \mu)^2$. This will focus the values around those of the global model, but also allow a moderate amount of variation.
This prior choice can be stated: 

\begin{equation}
  \label{eq:entrance_prior}
  \boldsymbol{\mathcal{U}}_{\textnormal{Exit}} \sim \mathcal{N}(\mu_{\textnormal{global}, k}, \{0.10 \, \mu_{\textnormal{global}, k}\}^2), 
\end{equation}

with the ``global" label referring to the values of Table~\ref{tab:opt_parms} and $k$ labeling each of the potential parameters for the exit channel.

At this point it is worth emphasizing that the potential priors for both the entrance and exit potentials are essentially arbitrary. The $20 \%$ and $10 \%$ variation for the parameters are meant to make this computation tractable, since it is impossible with the limited amount of data to uniquely determine the parameters as discussed in Sec.~\ref{sec:amb_pots}. The influence of this choice on the entrance channel is limited since there is data to inform the parameters. However, the choice of $10 \%$ for the exit channel will influence our final calculated uncertainties. Lower or higher amounts of variation could be considered for these parameters, but a choice has to be made in order to account for their impact on DWBA calculations. We have also chosen to excluded variations in the spin-orbit and bound state potentials. However, the possible impact of the bound state potentials will be discussed in Sec.~\ref{sec:discussion}.       
 
Our model treats $C^2S$ as another parameter to be estimated, so a prior must be specified. We have assigned it the mildly informative prior:
\begin{equation}
  \label{eq:cs_prior}
  C^2S \sim \textnormal{HalfNorm}(n_{nucleon}^2),  
\end{equation}

where $n_{nucleon}$ is the number of nuclei occupying the orbital that is involved in the transfer. The half-normal distribution ensures that $C^2S \geq 0$, while the scale parameter comes from the sum rules of Macfarlane and French \cite{sum_rules}. These rules have been found experimentally to be a robust constraint \cite{sum_rule_test}. However, it is likely that this prior is more conservative than necessary, since we do not expect a single state to contain the entirety of the strength for a given shell, but it serves as a rough estimate to help construct the prior for $C^2S$.

The use of the zero-range approximation for the transfer channels also comes with an additional uncertainty from the strength parameter, $D_0$, as discussed in
Sec.~\ref{sec:cs_calc}. Our model explicitly accounts for this $15 \%$ uncertainty by using a parameter $\delta D_0^2$, which is assigned a normal and informative prior:

\begin{equation}
  \label{eq:d0}
  \delta D_0^2 \sim \mathcal{N}(1.0, 0.15^2).
\end{equation}

We also introduced two additional parameters that are not a part of DWBA, but are instead meant to account for deficiencies in the reaction theory. The first is a normalization parameter, $\eta$, which allows for the adjustment of the theoretical predictions for both the elastic and transfer cross sections based on any observed normalization difference between the elastic channel data and optical model calculations. This can be in principle seen as treating the absolute scale of the data as arbitrary, which prevents biasing the potential parameters towards
unphysical values if a systematic difference is present. The posterior for this parameter will only be informed by the elastic data of the entrance channel, but will directly influence the posterior for $C^2S$. Since $\eta$ is multiplicative in nature, we do not want to bias it towards values less than or greater than $1$. This is done by introducing a
parameter, $g$, which is uniformly distributed according to:

\begin{equation}
  \label{eq:g_uni}
  g \sim \textnormal{Uniform}(-1, 1).
\end{equation}

\noindent $\eta$ is then defined as:

\begin{equation}
  \label{eq:eta}
  \eta = 10^{g}.
\end{equation}

\noindent Collecting all of these factors, we can now write the DWBA predictions at each angle $i$ as:
\begin{equation}
  \label{eq:cs_full}
  \frac{d \sigma}{d \Omega}^{\prime}_{\textnormal{DWBA}, i} = \eta \times \delta D_0^2 \times C^2S \times \frac{d \sigma}{d \Omega}_{\textnormal{DWBA}, i}.
\end{equation}

The second additional parameter comes from the consideration that the DWBA theory provides only an approximation to the true transfer cross section. If we only consider the measured experimental uncertainties
from the transfer channel, any deviation from DWBA will significantly influence the posteriors for the potential parameters. This is remedied by introducing an additional theoretical uncertainty, $\sigma_{\textnormal{theory}, i}$, where the index $i$ references the angle at which the differential cross section is evaluated. We estimate this quantity as a percentage uncertainty on the
theoretical cross section, which is based on a single unknown parameter, $f$.  Our total uncertainty at an angle is thus:

\begin{equation}
  \label{eq:unc}
  \sigma_i^{\prime 2} = \sigma_{\textnormal{Transfer}, i}^2 +  (f\frac{d \sigma}{d \Omega}^{\prime}_{\textnormal{DWBA}, i})^2.
\end{equation}

\noindent We use $\frac{d \sigma}{d \Omega}^{\prime}_{\textnormal{DWBA}, i}$ as defined in Eq.~(\ref{eq:cs_full}), $\sigma_{\textnormal{Transfer}, i}^2$ is the experimental statistical uncertainty, and the adjusted uncertainty, $\sigma_i^{\prime 2}$,
assumes that the experimental and theoretical uncertainties are independent. Since $f$ is some fractional amount of the predicted cross section, we assign it the weakly informative prior:

\begin{equation}
  \label{eq:f}
  f \sim \textnormal{HalfNorm}(1),
\end{equation}

\noindent so that it is biased towards values less than $1$.  

Finally, the likelihood functions for the experimental data must also be specified. The analysis of each excited state will
require two likelihood functions for both the elastic and transfer data. These likelihood functions use the normal distribution, and
take the form:

\begin{equation}
  \label{eq:likelihood}
  \frac{d \sigma}{d \Omega}_{\textnormal{Exp}, i} \sim \mathcal{N}(\frac{d \sigma}{d \Omega}_{\textnormal{Theory}, i}, \sigma_{\textnormal{Exp}, i}^2), 
\end{equation}

where $i$ again refers to a specific angle. This expression assumes that the residuals between the experimental cross section and the ones
calculated from theory are distributed normally. 

Taking into account all of the considerations and definitions listed above, we can write down our full Bayesian model.
Experimental elastic scattering data is identified by the label $\textnormal{Elastic}$, and the transfer data is labeled $\textnormal{Transfer}$. The theoretical differential cross sections calculated with FRESCO are written $\frac{d \sigma}{d \Omega}_{\textnormal{Optical}, j}$ for elastic scattering and $\frac{d \sigma}{d \Omega}_{\textnormal{DWBA}, i}$ for the transfer reaction. The indices $i$ and $j$ refer to the transfer and elastic angles, respectively. The model is, thus:   

\begin{align}
  \label{eq:model}
 & \textnormal{Priors:} \nonumber \\
 & \boldsymbol{\mathcal{U}}_{\textnormal{Entrance}} \sim \mathcal{N}(\mu_{\textnormal{central}, k}, \{0.20 \, \mu_{\textnormal{central}, k}\}^2) \nonumber \\
 & \boldsymbol{\mathcal{U}}_{\textnormal{Exit}} \sim \mathcal{N}(\mu_{\textnormal{global}, k}, \{0.10 \, \mu_{\textnormal{global}, k}\}^2) \nonumber \\
 & f \sim \textnormal{HalfNorm}(1) \nonumber \\
 & \delta D_0^2 \sim \mathcal{N}(1.0, 0.15^2) \nonumber \\
 & C^2S \sim \textnormal{HalfNorm}(n_{nucleon}^2) \nonumber \\
 & g \sim \textnormal{Uniform}(-1, 1) \nonumber \\
 & \textnormal{Functions:} \\
 & \eta = 10^{g}  \nonumber \\
 & \frac{d \sigma}{d \Omega}^{\prime}_{\textnormal{Optical}, j} = \eta \times \frac{d \sigma}{d \Omega}_{\textnormal{Optical}, j} \nonumber \\
 & \frac{d \sigma}{d \Omega}^{\prime}_{\textnormal{DWBA}, i} = \eta \times \delta D_0^2 \times C^2S \times \frac{d \sigma}{d \Omega}_{\textnormal{DWBA}, i} \nonumber \\
 & \sigma_i^{\prime 2} = \sigma_{\textnormal{Transfer}, i}^2 +  (f\frac{d \sigma}{d \Omega}^{\prime}_{\textnormal{DWBA}, i})^2 \nonumber \\
 & \textnormal{Likelihoods:} \nonumber \\
 & \frac{d \sigma}{d \Omega}_{\textnormal{Transfer}, i} \sim \mathcal{N}(\frac{d \sigma}{d \Omega}^{\prime}_{\textnormal{DWBA}, i}, \sigma_i^{\prime \, 2}) ,  \nonumber \\
 & \frac{d \sigma}{d \Omega}_{\textnormal{Elastic}, j} \sim \mathcal{N}(\frac{d \sigma}{d \Omega}^{\prime}_{\textnormal{Optical}, j}, \sigma_{\textnormal{Elastic}, j}^2) ,  \nonumber 
\end{align}

where the $k$ index runs over each of the potential parameters.

It should also be noted that the applicability of DWBA requires that the reaction is dominated
by a direct reaction mechanism occurring at the nuclear surface. Thus, transfer data must be collected at intermediate laboratory energies to suppress the contributions of isolated resonances and low angles to ensure a surface dominated reaction. Failure to adhere to these principles could introduce additional uncertainties into the extraction of $C^2S$. Practically, this work follows the suggestion of Ref. \cite{thompson_nunes_2009} and only fits the transfer data up to the first observed minimum in the data.

\subsection{Posterior and Evidence Estimation}

\underline{M}arkov \underline{C}hain \underline{M}onte \underline{C}arlo (MCMC) algorithms are one of the most common ways to calculate posterior distributions \cite{mcmc_review}. However, it is clear from Eq.~(\ref{eq:model}) that our Bayesian model lives in a high dimensional space, which
presents a difficult challenge for all MCMC algorithms.
In particular, traditional Metropolis-Hastings samplers require tuning of the step proposals for each dimension. This problem 
is avoided with the Affine Invariant Ensemble sampler of Goodman and Weare \cite{ensemble_mcmc}.
This method uses an ensemble of random walkers to 
sample the posterior, and has been designed to perform well with linearly correlated
parameters. We use the Python package \texttt{emcee} to implement the algorithm \cite{emcee}.
Using this method with a stretch move requires only a single parameter, $a$, to be specified.
{A detailed explanation of this parameter can be found in Appendix B.
$a$ is fixed to its suggested value $a=2$ in this work} \cite{emcee}.
The posteriors for each state are estimated using an ensemble of $400$ walkers which take $> 4000$ steps.
{Burn in periods were found to take approximately $1000$ steps.}
Final parameter estimates are taken from the final $2000$ steps, which are then thinned by $50$ in order to
give $1.6 \times 10^4$ samples. {The autocorrelation in the samples before thinning was estimated to be roughly $400$ steps. $2000$ steps would then contain 5 autocorrelation lengths, with each length yielding one independent sample per walker. This means we draw $\approx 2000$ independent samples from the posterior ensuring that the statistical fluctuations of the sampling are negligible compared to the uncertainties in the posteriors. Thinning was only used to reduce the number of samples and thereby ease subsequent calculations such as the credibility intervals for the differential cross sections.}  

MCMC methods draw samples directly from the posterior distribution which allows parameter estimation, but
they do not allow a straightforward estimation of the evidence integral. 
The model selection necessary to assign $\ell$ values requires the calculation of
Eq.~(\ref{eq:model_prob}). Monte Carlo integration techniques solve the issue
of calculating $Z$, but essentially reverse the previous issue by placing
a diminished focus on the calculation of the posterior distributions. This means that
separate calculations have to be carried out for our two tasks of parameter estimation (spectroscopic factors)
and model selection ($\ell$ assignment). Our evidence calculation is carried out using the nested sampling
procedure introduced by Skilling \cite{skilling2006, skilling2004}, as
implemented in the \texttt{dynesty} Python package \cite{speagle2019dynesty}. A brief description of this
algorithm is given in Appendix C.

For this work all nested sampling runs used $250$ live points bounded with multiple ellipsoids and updates performed through slice sampling.
The stopping criteria was set at $\Delta Z_i < .01$. Since the nested sampling is subject to statistical uncertainties in $\ln Z$, it is necessary to propagate these uncertainties to both $B_{ij}$ and the probabilities for each $\ell$ transfer defined by Eq.~(\ref{eq:model_prob}). This was done by drawing $10^6$ random samples from the Gaussian distributions for each $\ln Z_i$, and then applying either Eq.~(\ref{eq:model_prob}) or Eq.~(\ref{eq:bayes_factor}) to each sample, yielding a set of samples for each quantity. From these samples we report the $68 \%$ credibility intervals, constructed from the $16$, $50$, and $84$ percentiles. 

\section{Analysis of $^{70} \textnormal{Zn}(d, ^{3} \textnormal{He})^{69} \textnormal{Cu}$}
\label{sec:results}

\begin{figure*}
  \centering
  \captionsetup[subfigure]{labelformat=empty}
    \vspace{-1\baselineskip}
    \begin{subfigure}[t]{0.45\textwidth}
        \includegraphics[width=\textwidth]{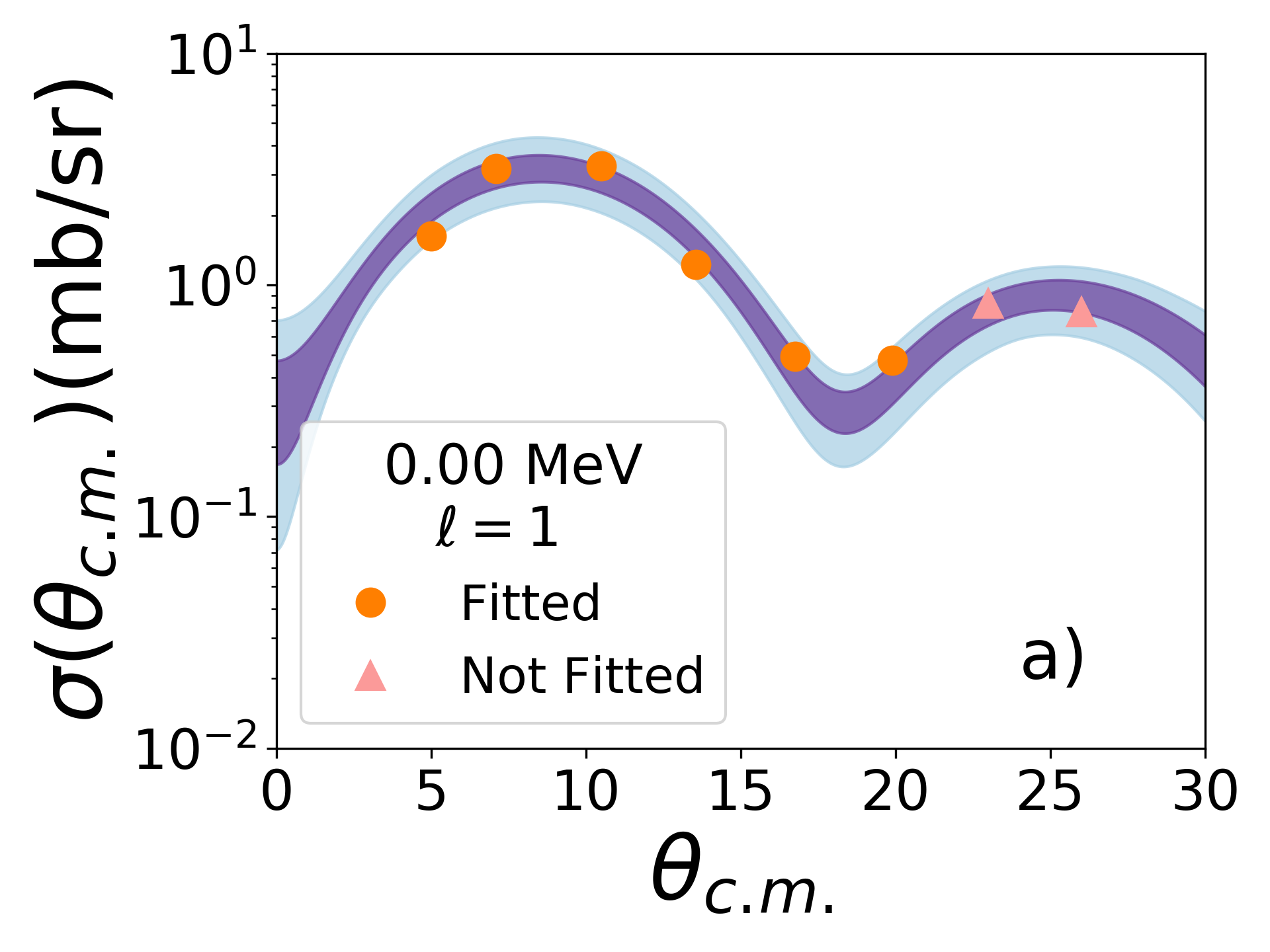}
        \caption{\label{fig:gs_fit}}
      \end{subfigure}
          \vspace{-1\baselineskip}
    \begin{subfigure}[t]{0.45\textwidth}
      \includegraphics[width=\textwidth]{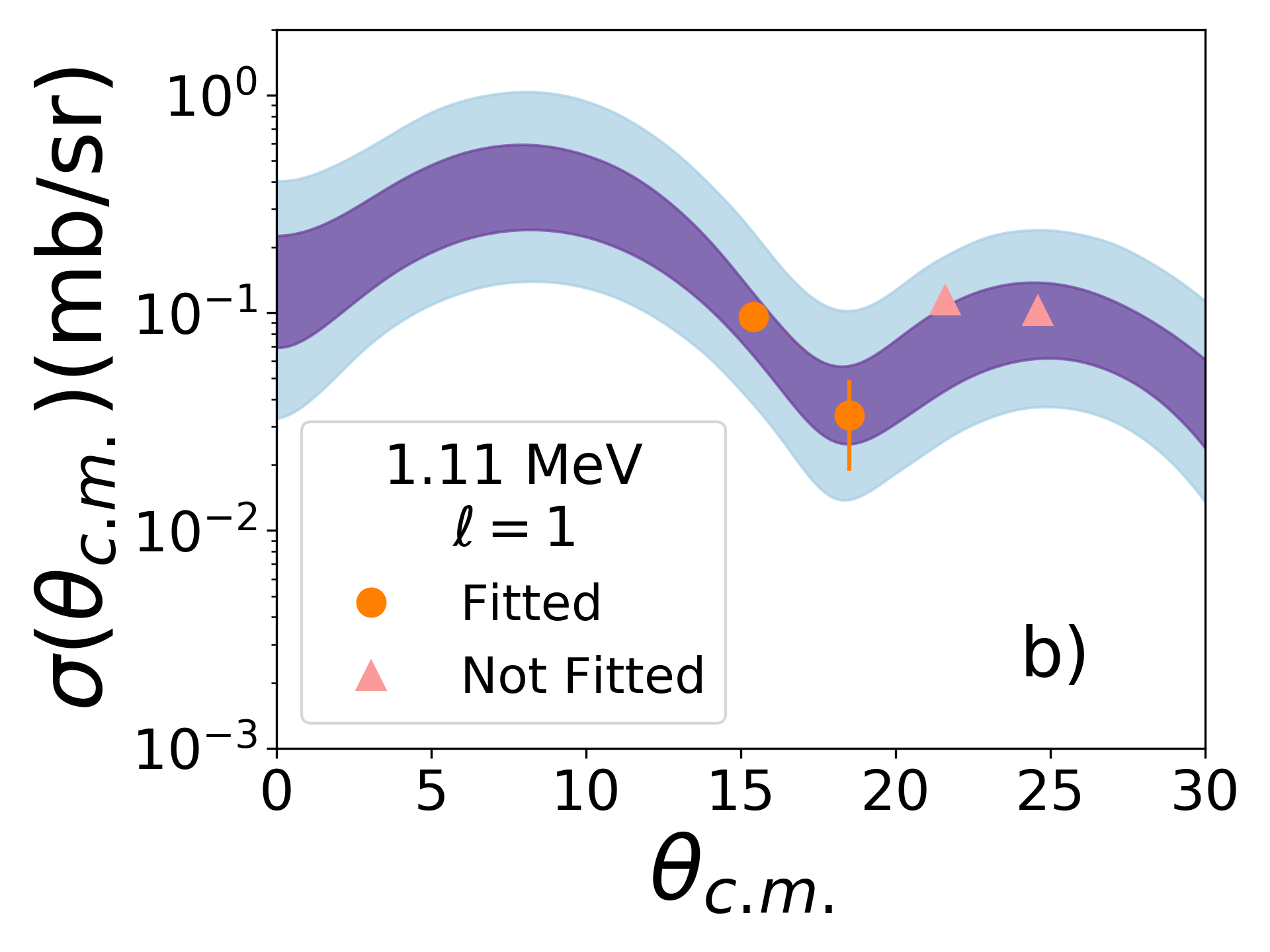}
          \vspace{-1\baselineskip}
      \caption{\label{fig:111_fit}}
    \end{subfigure}
    \vspace{-1\baselineskip}
    \begin{subfigure}[t]{0.45\textwidth}
      \includegraphics[width=\textwidth]{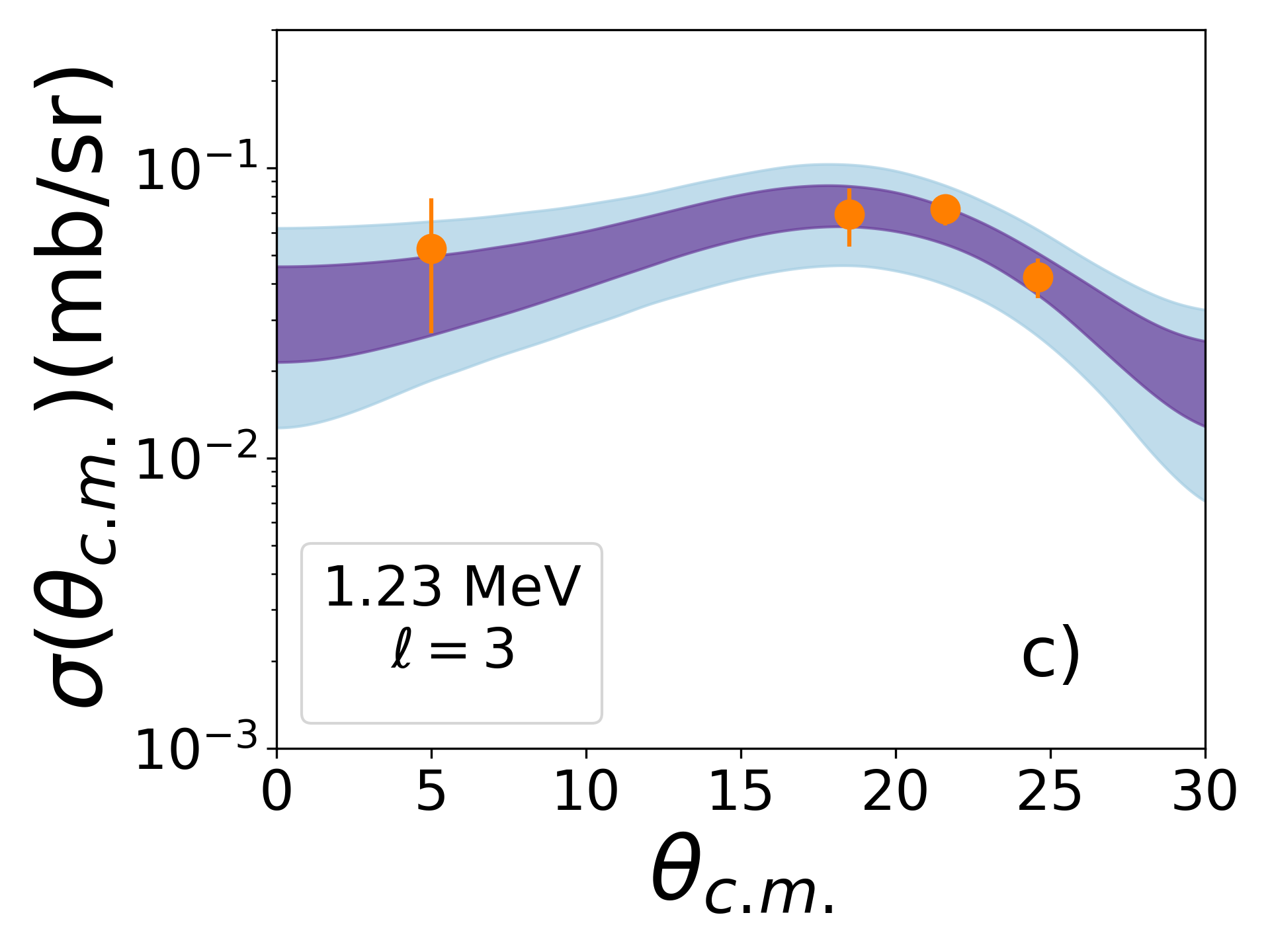}
          \vspace{-1\baselineskip}
      \caption{\label{fig:123_fit}}
    \end{subfigure}
    \begin{subfigure}[t]{0.45\textwidth}
      \includegraphics[width=\textwidth]{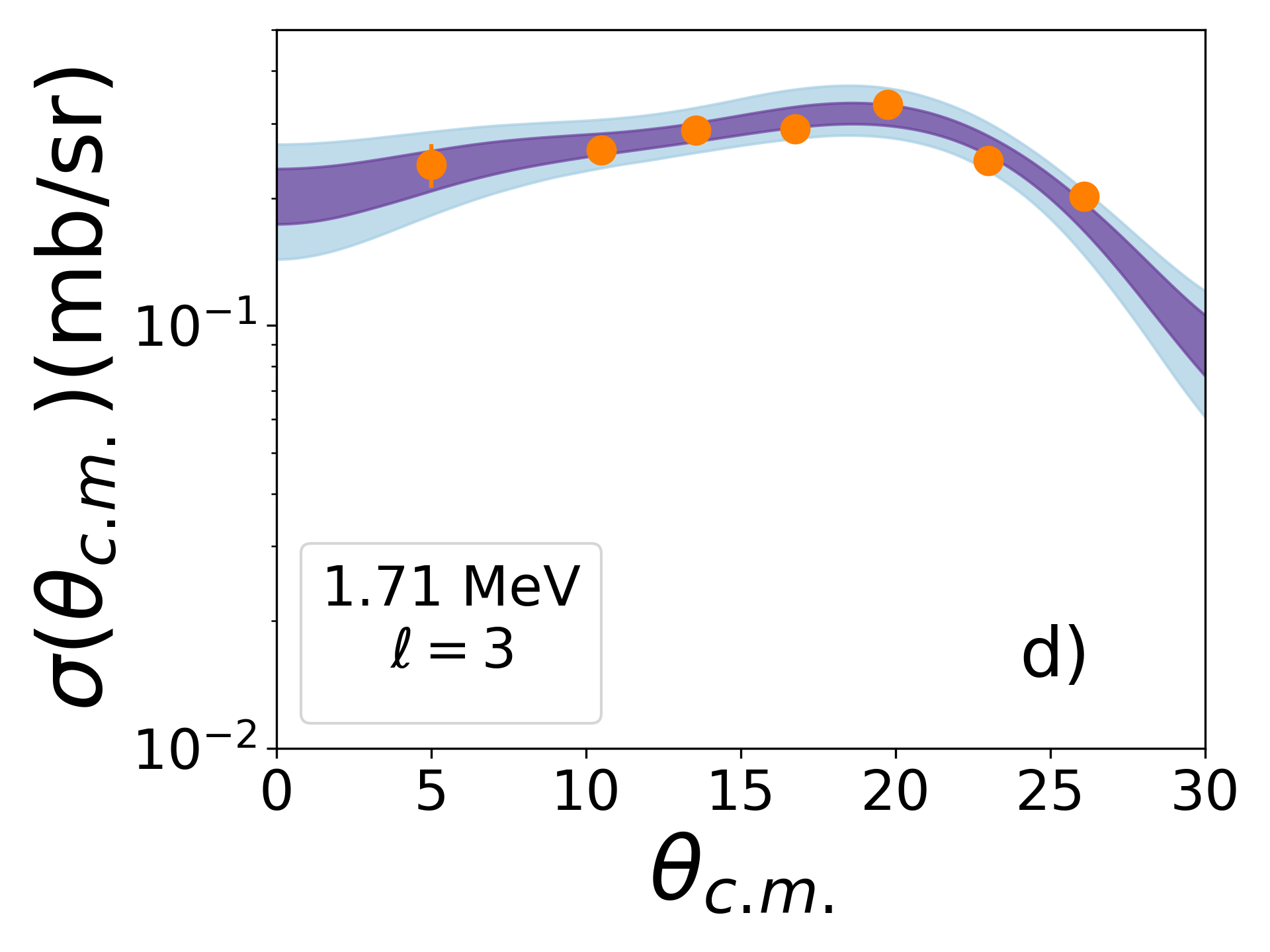}
          \vspace{-1\baselineskip}
      \caption{\label{fig:171_fit}}
    \end{subfigure}
    \vspace{-1\baselineskip}
    \begin{subfigure}[t]{0.45\textwidth}
      \includegraphics[width=\textwidth]{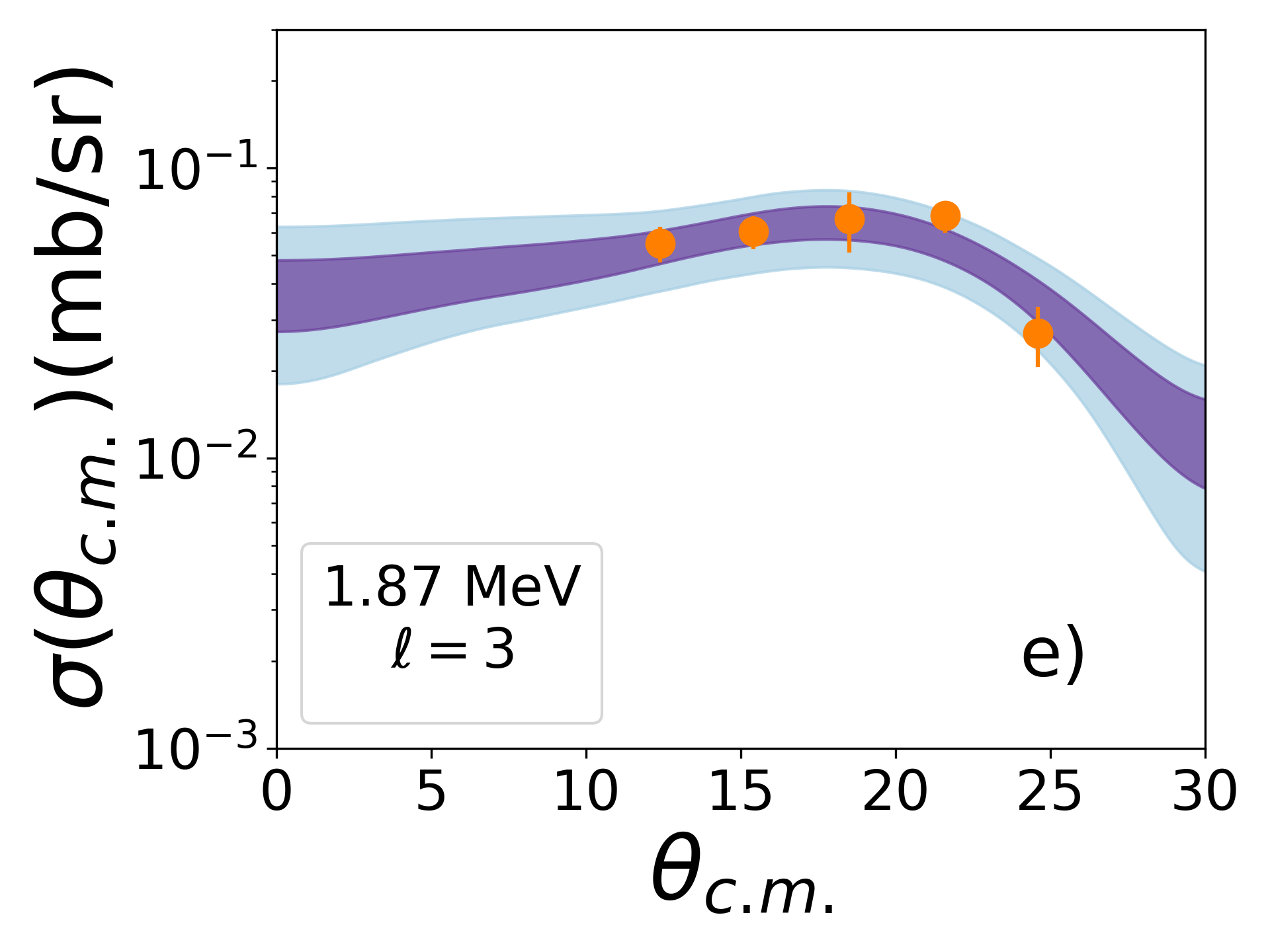}
          \vspace{-1\baselineskip}
      \caption{\label{fig:187_fit}}
    \end{subfigure}
    \begin{subfigure}[t]{0.45\textwidth}
      \includegraphics[width=\textwidth]{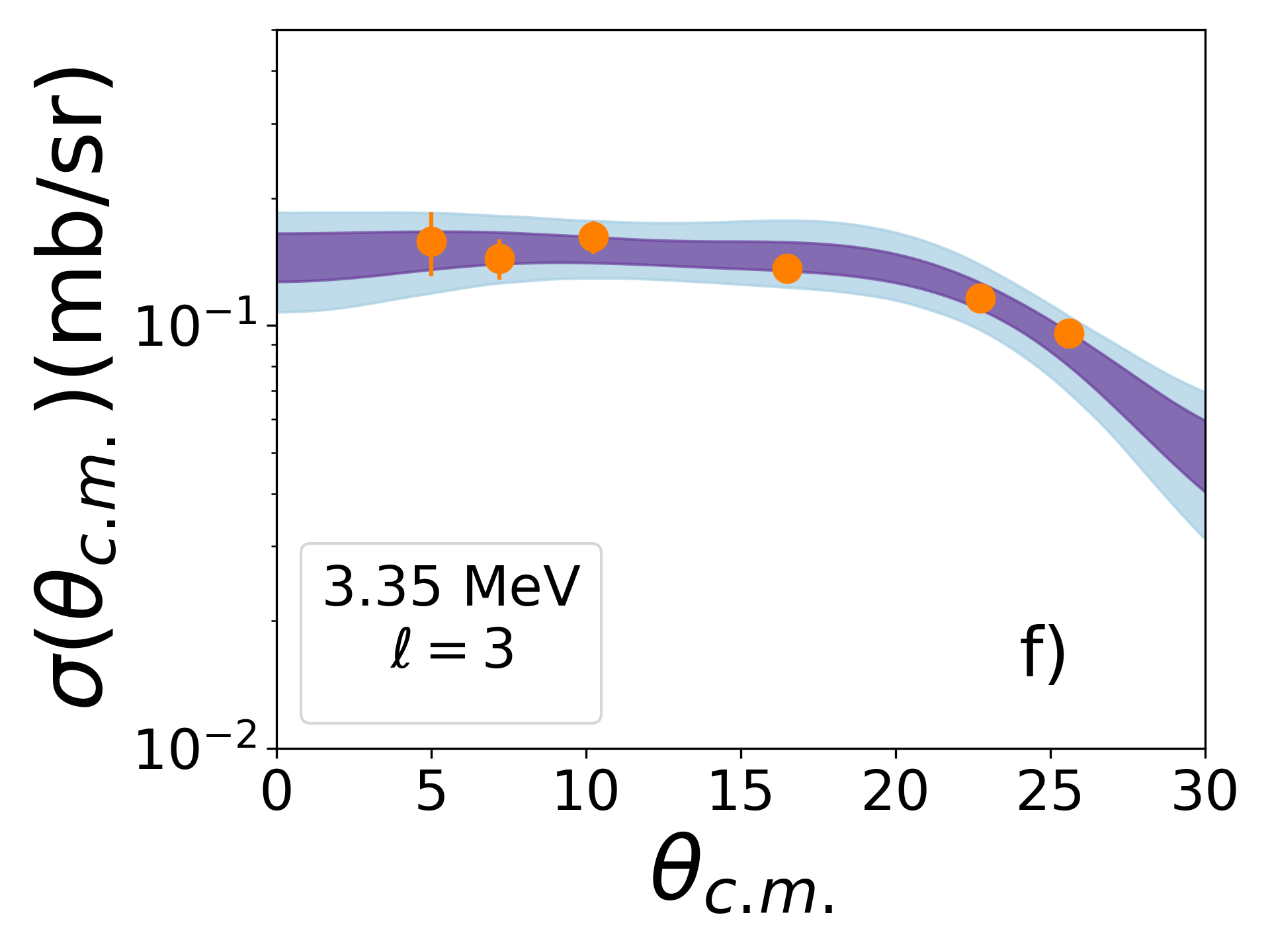}
          \vspace{-1\baselineskip}
      \caption{\label{fig:335_fit}}
    \end{subfigure}
    \vspace{-1\baselineskip}
    \begin{subfigure}[t]{0.45\textwidth}
      \includegraphics[width=\textwidth]{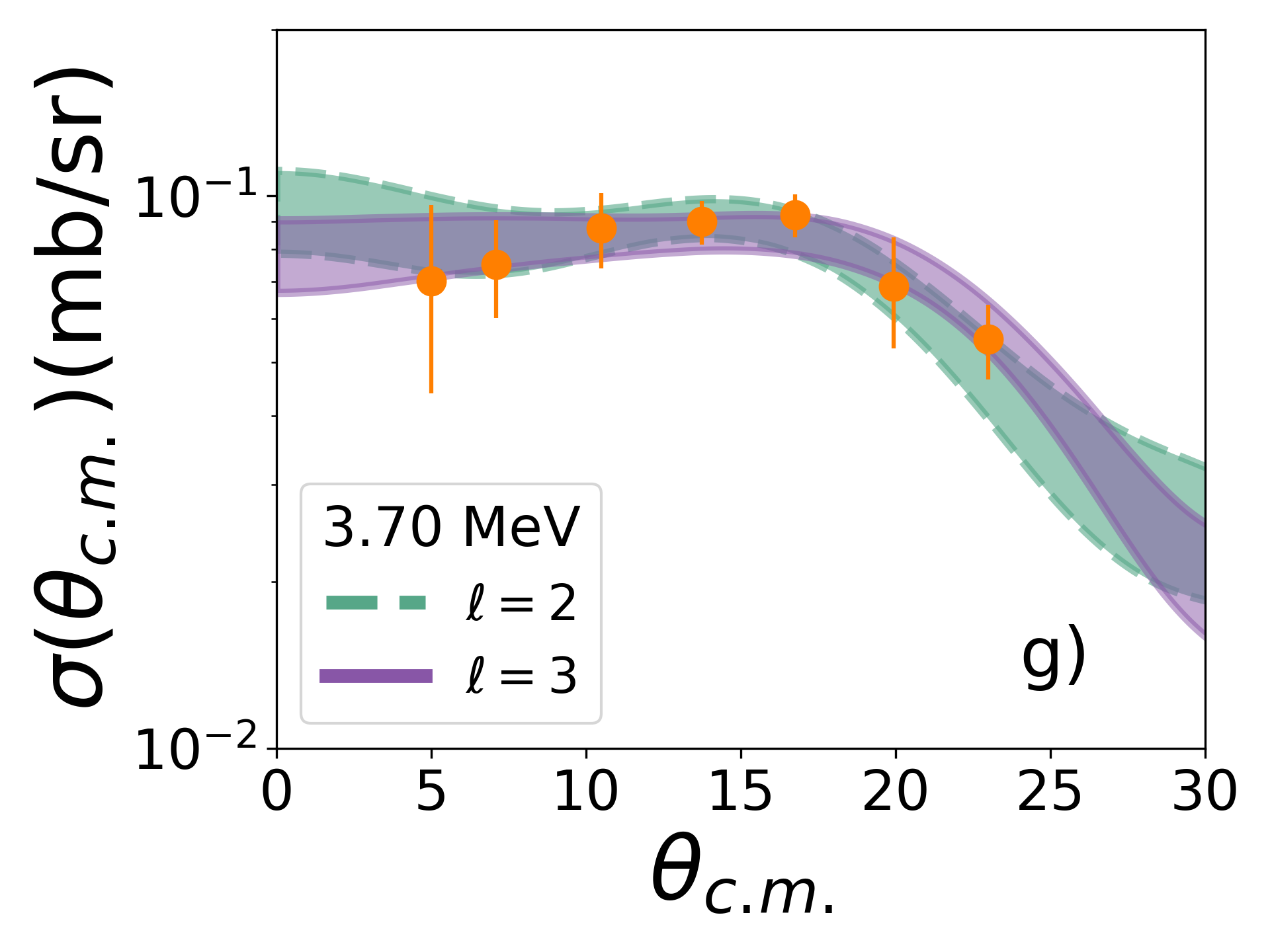}
          \vspace{-1\baselineskip}
      \caption{\label{fig:370_fit}}
    \end{subfigure}
    \begin{subfigure}[t]{0.45\textwidth}
      \includegraphics[width=\textwidth]{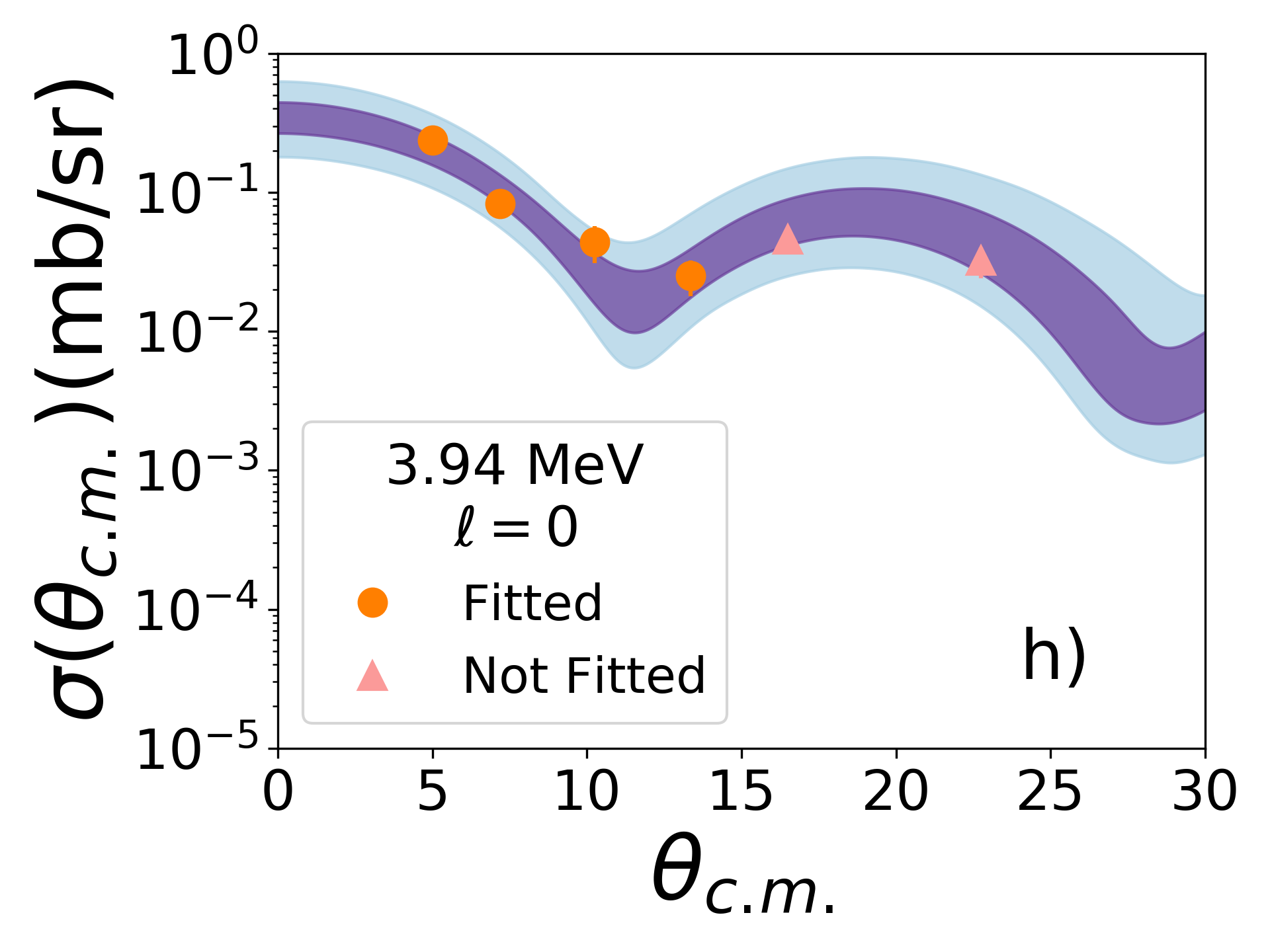}
          \vspace{-1\baselineskip}
      \caption{\label{fig:394_fit}}
    \end{subfigure}
    \vspace{-1.8\baselineskip}
    \caption{\label{fig:states} DWBA calculations for the states of $^{69}$Cu . The $68 \%$ and $95 \%$ credibility intervals are shown in purple and blue, respectively. Only the data points shown in orange circles were considered in each calculation. For the
      $3.70$ MeV state the $68 \%$ bands are shown for the two most likely $\ell$
    transfers.}
\end{figure*}

The Bayesian model of Sec.~\ref{sec:bayes_stuff} allows us to extract spectroscopic factors and assign $\ell$ values
to observed transfers, while taking into account uncertainties associated with the optical potentials.
In order to test these methods, we will focus on a reanalysis of the $^{70}$Zn$(d, ^{3}$He$)^{69}$Cu reaction,
which was originally presented in Ref.~\cite{pierre_paper}. For reference, data was collected by impinging a $27$ MeV deuteron beam onto
a thin target of enriched $^{70}$Zn. The reaction products were measured with a magnetic spectrograph. The original study should be referred to for complete experimental details. This reaction and the measured data set have two important conditions that simplify our study. First, since $^{70} \textnormal{Zn}$ has a $0^{+}$ ground state, only a unique $\ell$ transfer is allowed for a given final state. Second, only 8 low lying bounds states were observed, meaning no additional
theoretical model is needed for treating transfers to the continuum. Our results
are summarized in Table~\ref{tab:cs}. Comparisons are made to the original values of the zero-range {and finite-range} calculations of the previous work.
Plots of the DWBA cross sections generated from the MCMC calculations are shown in Fig.~\ref{fig:states}. The purple and blue bands show the $68 \%$ and $95 \%$ credibility bands, respectively. Using samples directly from the Markov chain means that these credibility bands accurately account for all of the correlations present between the parameters. Each of these states will now be discussed in detail, with additional calculation details provided for the ground state in order to demonstrate
the use of our Bayesian method.

\begin{table*}[ht]
  \begin{threeparttable}
  \setlength{\tabcolsep}{4pt}
  \caption{\label{tab:cs} Summary of the spectroscopic factors derived in this work. {Comparisons to the zero-range (ZR) and finite-range (FR) calculations of} Ref.~\cite{pierre_paper} are made. {All calculations use the same bound state parameters}.}
  \begin{tabular}{cccccc}
    \addlinespace[0.5mm]
    \\ \hline \hline
    \\ [-1.0 ex]
    $E_x$(MeV) & $\ell$ & $J^{\pi}$ \tnote{a} & $C^2S(ZR)$ \cite{pierre_paper} &  $C^2S(FR)$ \cite{pierre_paper}   & $C^2S$(This work)           \\ [-.5ex] 
    \\ \hline 
    \\ [-1.5ex] 
$0.0$      & $1$    & $3/2^{-}$ & $1.40(15)$  & $1.50(17)$  & $2.06^{+0.87}_{-0.68}$            \\ [0.8ex]
$1.11$     & $1$    & $1/2^{-}$ &    -        & $0.35(11)$  & $0.48^{+0.52}_{-0.25}$            \\ [0.8ex]
$1.23$     & $3$    & $(5/2^{-})$ & $0.80(11)$& $0.70(10)$  & $1.10^{+0.81}_{-0.48}$   \\ [0.8ex]
$1.71$     & $3$    & $7/2^{-}$ & $2.00(11)$  & $2.50(14)$  & $2.37^{+1.36}_{-0.84}$            \\ [0.8ex]
$1.87$     & $3$    & $7/2^{-}$ & $0.40(10)$  & $0.50(10)$  & $1.07^{+0.93}_{-0.51}$            \\ [0.8ex]
$3.35$     & $3$    & $(7/2^{-})$ & $1.60(10)$ & $2.40(15)$  & $2.67^{+1.83}_{-1.06}$             \\ [0.8ex]
\multirow{2}{*}{$3.70$}     & $2$    & $(3/2^{+})$ &  $1.90(25)$ & $1.50(20)$  & $1.74^{+1.05}_{-0.62}$             \\ [0.8ex]
           & $3$    & $(7/2^{-})$ & -            & - & $2.90^{+2.75}_{-1.43}$             \\ [0.8ex]
    $3.94$     & $0$    & $1/2^{+}$ & $0.70(6)$    & $0.70(10)$  & $1.03^{+0.71}_{-0.44}$          \\ [-1.5ex]
    \\ \hline \hline
  \end{tabular}
  \begin{tablenotes}
\item[a] These assignments are discussed in depth in Sec.~\ref{sec:gs_sec} through Sec.~\ref{sec:394_sec}.
\end{tablenotes}
\end{threeparttable}
\end{table*}

\subsection{Ground State}
\label{sec:gs_sec}

\begin{figure}
  \centering
  \includegraphics[width=.4\textwidth]{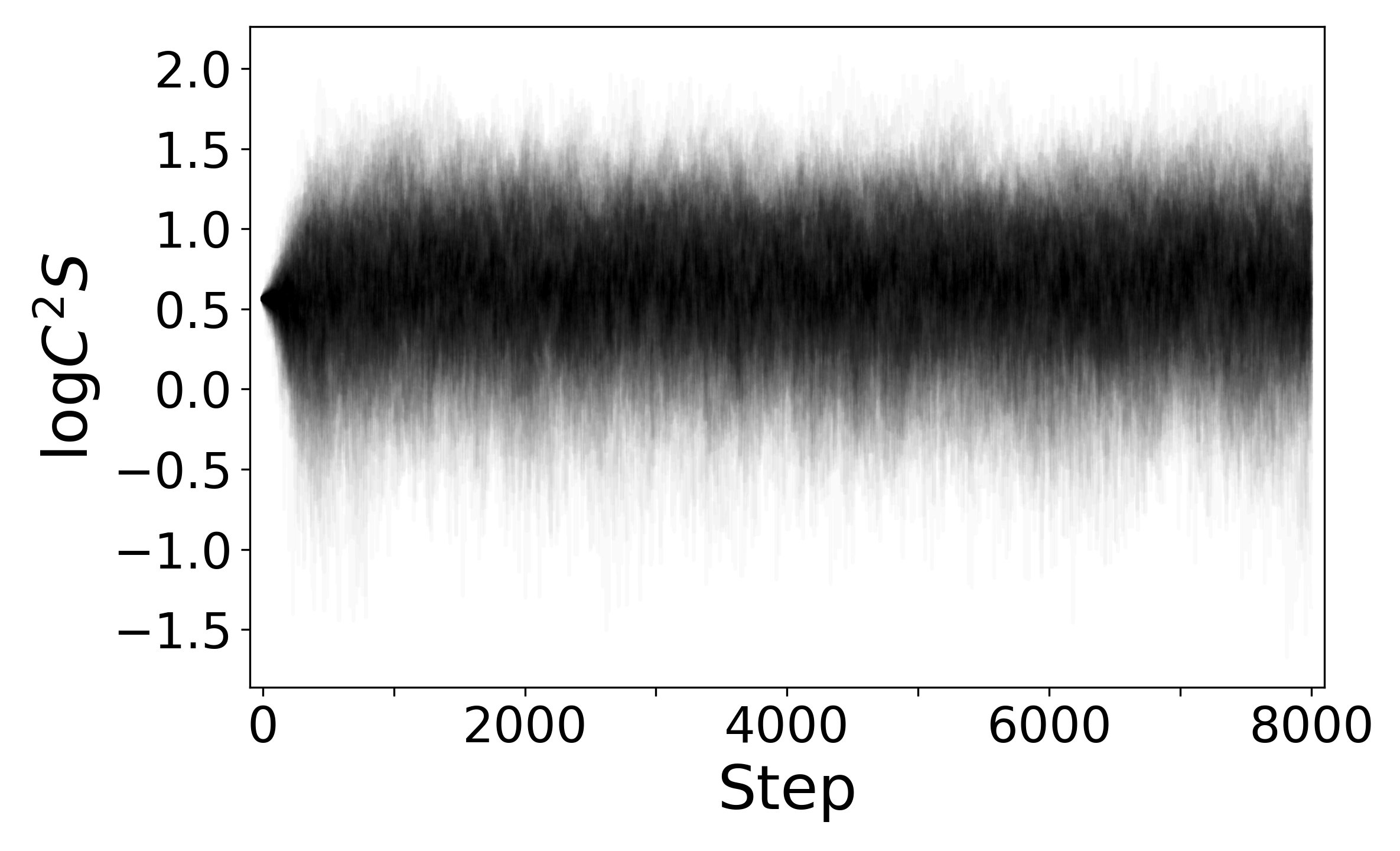}
  \caption{Trace of the MCMC walkers as a function of step and $\log(C^2S)$.
    Only the last 2000 steps were used for the posteriors.}
  \label{fig:steps}
\end{figure}

The MCMC calculations for the ground state were carried out using $8000$
steps and $400$ walkers in the ensemble. As an example we have provided the
trace plot for the value of $C^2S$ as a function of step in Fig.~\ref{fig:steps}.
Parameter values were estimated by using the last $2000$ steps and thinning by $50$.   

\begin{figure}
  \centering
  \includegraphics[width=.4\textwidth]{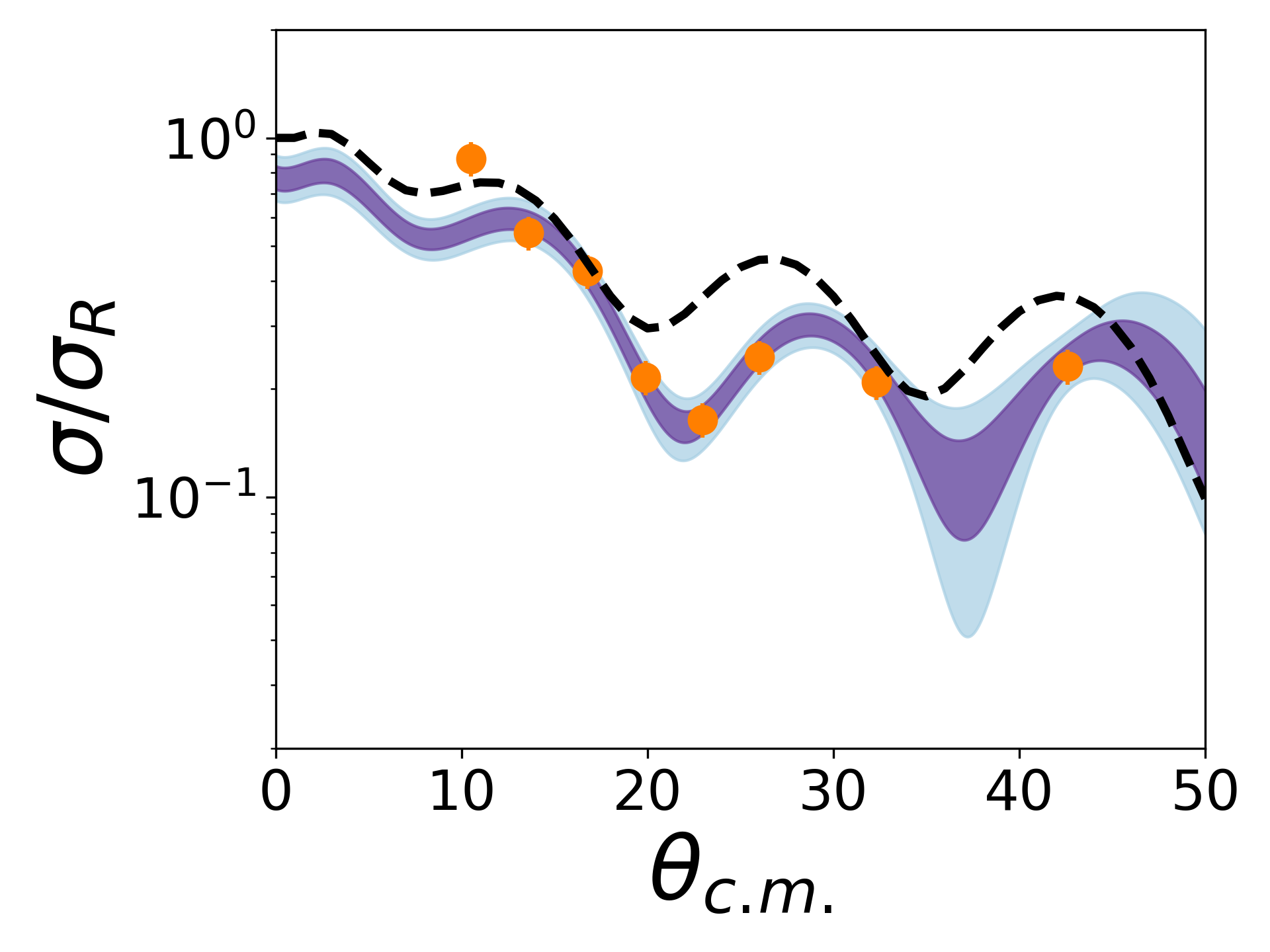}
  \caption{\label{fig:elastic_mcmc} Bayesian fit of the elastic data calculated simultaneously with the $0.00$ MeV state. The $68 \%$ and $95 \%$ credibility intervals are shown in purple and blue respectively, {while the black dashed curve was calculated using the global values from Table~\ref{tab:opt_parms}}.}
\end{figure}

\begin{figure*}
  \centering
  \includegraphics[width=\textwidth]{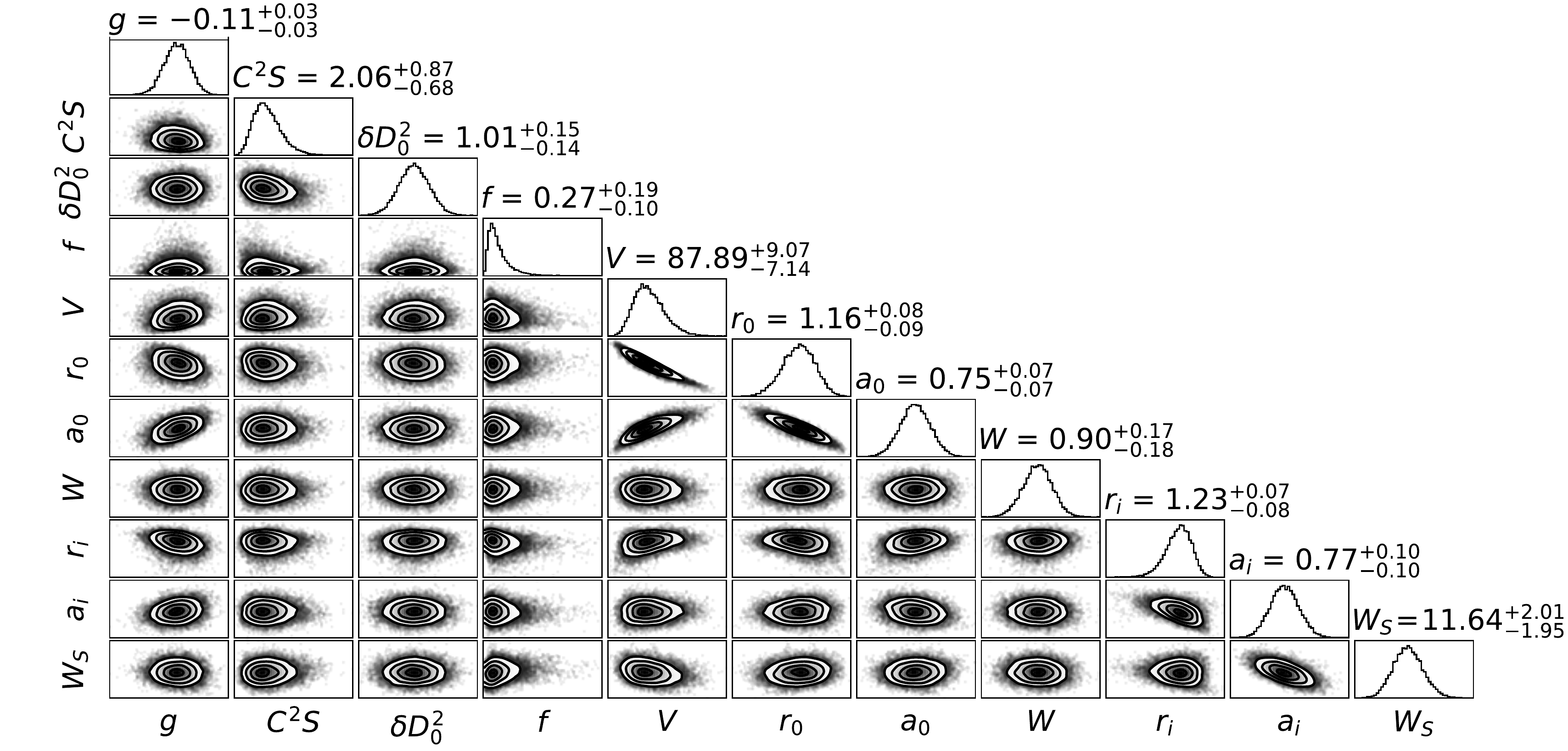}
  \includegraphics[width=\textwidth]{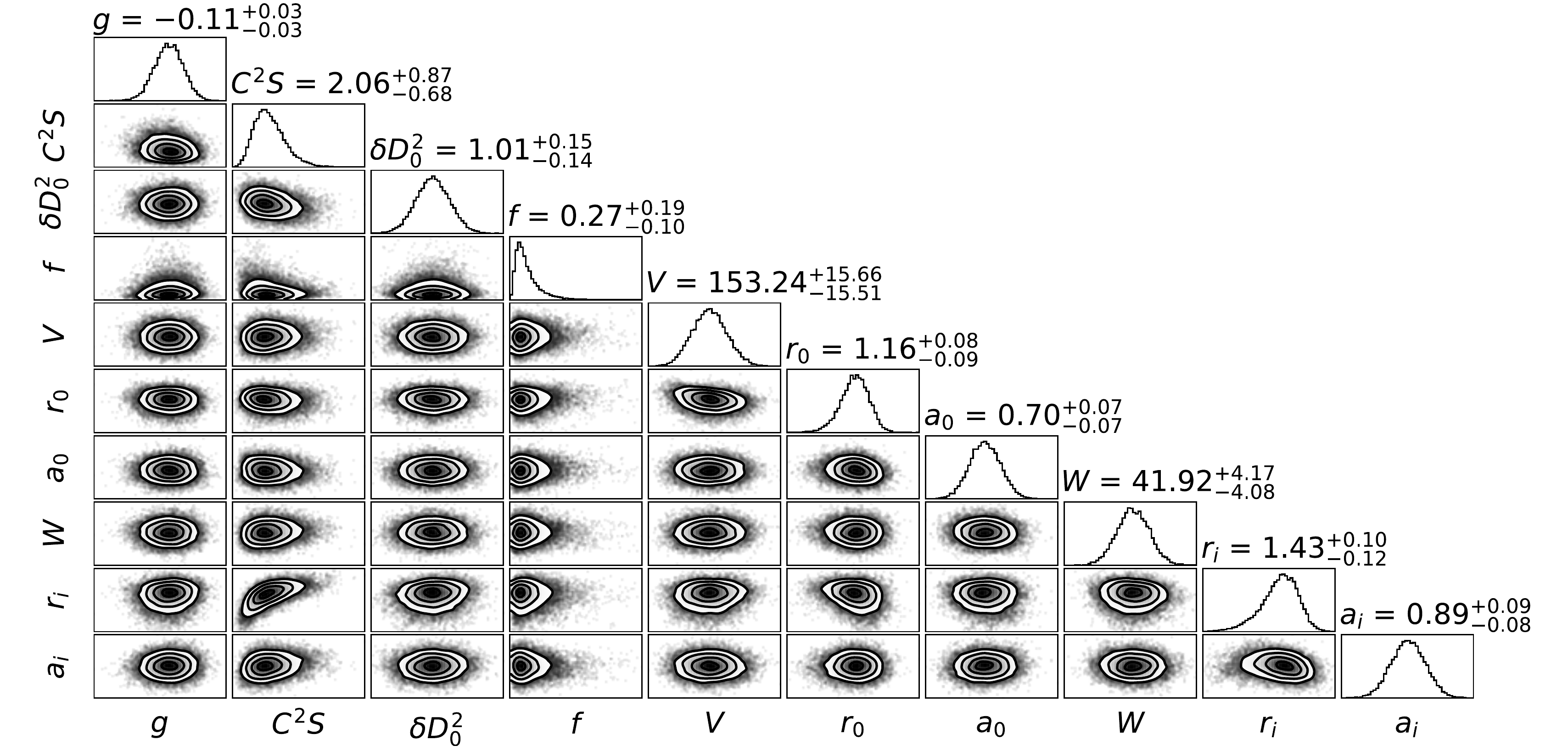}
  \caption{\label{fig:gs_corner} The pair-wise correlation plots for the ground state transfer. The top plot shows the entrance potential
    parameters, while the bottom shows the exit channel parameters. Both channels are compared to the model parameters as defined in Eq.~(\ref{eq:model}).
    The $68 \%$ credibility intervals are listed at the top of each column with the dashed lines showing their relationship to the $1$-D parameter distributions.
  The plots were generated using the Python package \texttt{corner} \cite{Foreman-Mackey2016}.}
  
\end{figure*}

As noted before, all of our MCMC calculations simultaneously fit the elastic scattering and transfer data. {This means that the posterior distributions shown in Fig.~\ref{fig:gs_corner} are functions of both the elastic and ground state transfer data.} The impacts of
the choice of potential parameters and the scale parameter $\eta$ on the elastic fit is quite dramatic. If we were
to adopt the global values in Table~\ref{tab:opt_parms} without adjusting any parameters, the agreement between theory and experiment would be quite poor as shown by the dashed black line in Fig.~\ref{fig:elastic_mcmc}. It should also be noted that the experimental
uncertainties for these points are roughly $10 \%$. On the other hand, the purple and blue bands in Fig.~\ref{fig:elastic_mcmc} show the fit obtained when we use our Bayesian model, which quite clearly provides a better description of the data. A significant difference is found between the normalization of the data and the optical model prediction, with $\eta \simeq 23 \% $.

By examining the correlations between the parameters, our model should display the continuous ambiguity discussed in Sec.~\ref{sec:amb_pots}. The pair-wise correlation plots in Fig.~\ref{fig:gs_corner} show the posterior samples from the entrance (top) and exit (bottom) channel potentials and how they relate to those of $g$, $C^2S$, $\delta D_0$, and $f$. The intra-potential correlations are quite striking for the entrance channel. All of the real potential parameters, $V, r_0,$ and $a_0$, show strong correlations with one another, and slightly weaker correlations existing between $V$, $r_0$, $r_i$, and $W_s$. Strong relationships also exist between $a_i$ and $W_s$, which is also another known continuous ambiguity \cite{perey_perey}. There is a much different situation for the exit potentials, where almost no intra-potential correlations
are seen. This result is expected since there are no elastic scattering data to constrain these parameters and because the Bayesian model parameter $f$ limits the amount of information that can be drawn from the transfer channel data. However, there is a surprisingly strong relationship between the exit channel imaginary radius and $C^2S$. A similar relationship can be seen with the entrance channel imaginary radius, but the effects on $C^2S$ are dramatically less.     

The results of the fit for the ground state are shown in Fig.~\ref{fig:gs_fit}.
The circular orange data points were the only data considered in the fit in order to
not bias our deduced spectroscopic factor as discussed in Sec.~\ref{sec:model}.  
The ground state of $^{69}$Cu is known to have a spin-parity of $\frac{3}{2}^-$, so the
transfer was calculated assuming a $2p_{3/2}$ state.


\subsection{1.11 MeV State}

The $1.11$ MeV state was only seen at four angles. Furthermore, only the first two data points lie within the
first minimum. The $J^{\pi} = \frac{1}{2}^-$ assignment is based on the observed angular distributions of Ref.~\cite{69cu_orig} and the analyzing power measurement of Ref.~\cite{fay_69cu}. In order to check that the data analyzed in the current work are consistent with these conclusions, the evidence integrals were calculated for $\ell = 0, 1, 2$, and $3$  transfers using all the data points.
The data support an $\ell = 1$ transfer, but do not rule out an $\ell = 3$ transfer. For this case, the median Bayes Factor defined in Eq.~(\ref{eq:bayes_factor}) is $B_{13} = 6.32$ (i.e the fifty percentile of $Z_1/Z_3$), indicating that there is substantial evidence in favor of $\ell = 1$. Since the data are consistent with the $\ell$ assignments of Ref.~\cite{fay_69cu,fay_69cu}, we carried out the MCMC calculations assuming a $2p_{1/2}$ state. The results of this calculation are plotted in Fig.~\ref{fig:111_fit}.      


\subsection{1.23 MeV State}

The state located at $1.23$ MeV is definitely associated with an $\ell = 3$
transfer. The previous analysis assumed a firm $J^{\pi} = \frac{5}{2}^-$; however,
the literature does not provide direct evidence for this. The analyzing power of Ref.~\cite{fay_69cu} was inconclusive, and the authors suggested the presence of a doublet based on the observed width of the peak in the spectrum. The $(d, ^3 \textnormal{He})$ experiment of Ref.~\cite{69cu_orig} also suggested a doublet and noted the high spectroscopic factor obtained ($C^2S = 1.5$) if a $\frac{5}{2}^-$ assignment was assumed. Other studies have also assigned a firm $J^{\pi} = \frac{5}{2}^-$ \cite{franchoo_mono, coul_ex, beta_decay}, but it is unclear if these results are actually independent determinations, or if they follow Table II of Ref.~\cite{69cu_orig}. We therefore follow the ENSDF evaluation \cite{a_69_ensdf} by recommending $J^{\pi} = (\frac{5}{2}^-, \frac{7}{2}^-)$, but only present the $C^2S$ value for $1f_{5/2}$ with the fit shown in Fig.~\ref{fig:123_fit}.


\subsection{1.71  and 1.87 MeV States}

From the parity constraints of Ref.~\cite{69cu_orig, fay_69cu} and the $\gamma$-ray anisotropies observed in Ref.~\cite{deep_inelas}, a firm $J^{\pi} = \frac{7}{2}^-$ assignment has been made for the $1.71$ MeV state. The results from our DWBA fit for a $1f_{7/2}$ state are shown in Fig.~\ref{fig:171_fit}. The arguments from the $1.71$ MeV state also apply to the state at $1.87$ MeV. A firm 
$J^{\pi} = \frac{7}{2}^-$ was assumed and a fit for a $1f_{7/2}$ state is shown in Fig.~\ref{fig:187_fit}.


\subsection{3.35 MeV State}

The state at $3.35$ MeV was reportedly seen in Ref. \cite{69cu_orig}, but no information was presented other than its possible existence. The previous analysis found an $\ell = 3$ nature to the angular distribution, and made a tentative assignment of $J^{\pi} = (\frac{7}{2}^-)$ assignment. Our methods support this conclusion as shown in Table~\ref{tab:probs}. $B_{3 \ell} > 10$ for each other $\ell$ transition, indicating strong evidence for the $\ell = 3$ transfer. The probability the final state was populated with an $\ell=3$ transfer is $P(\ell=3) = 91^{+3}_{-4} \%$. However, DWBA is still unable to discriminate between $J^{\pi} = (\frac{5}{2}^-, \frac{7}{2}^-)$. Our fit assuming a $1f_{7/2}$ state is shown in Fig.~\ref{fig:335_fit}. 

\begin{table*}[]
  \caption{\label{tab:probs} Results of the model comparison calculations for the $3.35$ and $3.70$ MeV states. For each $\ell$ value we list the $\log{Z}$ value calculated
with nested sampling, the median Bayes factor when compared to the most likely transfer $\ell=3$, and the probability of each transfer.}
\begin{tabular}{ccccc}
  \\ [0.5ex] \hline \hline
  \\ [-2.0ex]

  & $\ell$ & $\log{Z}_{\ell}$  & $B_{3 \ell}$ & $P(\ell)$         \\ [0.5ex] \hline
  \\ [-2.0ex]
\multirow{4}{*}{$E_x = 3.35 $ MeV} & 0      & 3.856(330)                                          & $> 10^4$     & $< .01 \%$        \\ [1.0ex]
                                   & 1      & 10.662(359)                                         & $15.94$      & $6^{+3}_{-2} \%$  \\ [1.0ex]
                                   & 2      & 9.961(363)                                          & $32.14$      & $3^{+2}_{-1} \%$  \\ [1.0ex]
  & 3      & 13.431(349)                                         & 1.0          & $91^{+3}_{-4} \%$ \\ 
  \\  [-5.0ex]
      \\  \hline 
      \\  [-2.0ex]
\multirow{4}{*}{$E_x = 3.70 $ MeV} & 0      & 10.393(365)                                         & $> 10^3$     & $ < 0.02 \%$      \\ [1.0ex]
                                   & 1      & 14.947(351)                                         & $45.98$      & $2^{+1}_{-1} \%$  \\ [1.0ex]
                                   & 2      & 16.640(346)                                         & $8.47$       & $10^{+5}_{-4} \%$ \\ [1.0ex]
                                   & 3      & 18.776(336)                                         & 1.0          & $88^{+4}_{-6} \%$ \\ [0.5ex] \hline \hline
\end{tabular}
\end{table*}




\subsection{3.70 MeV State}

\begin{figure}
  \centering
  \includegraphics[width=.4\textwidth]{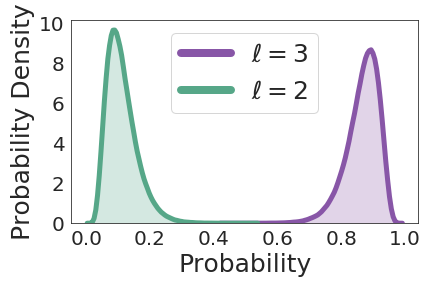}
  \caption{\label{fig:l_comp} The KDE representations of the probabilities of the $\ell=2,3$ transfers for the 3.70 MeV state.}
\end{figure}

The state at $3.70$ MeV was also seen for the first time in Ref.~\cite{pierre_paper}. However, our Bayesian method indicates an ambiguous $\ell$ assignment. As can
be seen in Fig.~\ref{fig:370_fit}, the measured angular distribution
is relatively flat, and does not appear to differ
from other states with $\ell = 3$. However, an assignment of $\ell = 2$ was
made in the previous analysis. Comparing the evidence integral for each case, we
indeed find the data effectively rule out $\ell = 0$ and $1$, while supporting an
$\ell = 2$ or $3$ assignment. Looking at Table~\ref{tab:probs}, we find a Bayes factor of $B_{32} = 8.47$ for
$\ell = 3$ over $\ell=2$, which suggests substantial evidence in favor of the $\ell = 3$ assignment. Using Eq.~(\ref{eq:model_prob}), the $68 \%$ credibility intervals for the probabilities are $P(\ell=3) = 88^{+4}_{-6} \%$ and  $P(\ell=2) = 10^{+5}_{-4} \%$, with the uncertainties coming from the statistical uncertainties of the nested sampling evidence estimation. The \underline{K}ernel \underline{D}ensiy \underline{E}stimates (KDE) for
the two dominate transfers are shown in Fig.~\ref{fig:l_comp} \cite{kde}. Our fits for both $\ell = 2$ and $3$ are shown if Fig.~\ref{fig:370_fit}.



\subsection{3.94 MeV State}
\label{sec:394_sec}

The $3.94$ MeV state was also observed for the first time in the previous study. The suggested $\ell = 0$ assignment was found to be supported by the data. The second most likely transfer was found to be $\ell = 1$. In this case $B_{01} = 72.24$, indicating very strong evidence in favor of the $\ell = 0$ assignment. The transfer to a $2s_{1/2}$ state is shown in Fig.~\ref{fig:394_fit}.

\section{Discussion}
\label{sec:discussion}

\subsection{Spectroscopic Factors}

The results of the previous section merit closer examination, especially with regards to the spectroscopic factors. Comparing our results with
those previously obtained in Table~\ref{tab:cs}, two things are clear: our median values tend to be larger then Ref.~\cite{pierre_paper} and
the uncertainties are much larger. To the first point, a majority of the shift comes from the lower value of $W_s$ used in the previous analysis.
Though not stated in Ref.~\cite{pierre_paper}, the surface potential was given a value of $W_s \approx 7.5$ MeV, which has the effect of lowering the value of $C^2S$.
Our values on average are higher due to the Bayesian analysis favoring $W_s = 11.93$ MeV and the inclusion of $\eta$, but these are somewhat offset due to the posterior values of $r_i$ and $a_i$ being lower than their global values. To the second point, when all of the sources of uncertainty are included in the analysis, we find highly asymmetric and data-driven uncertainties on $C^2S$ ranging from $35-108 \%$. This is a substantial increase with regards to the common assumption that the extraction of spectroscopic factors comes with an approximately $25 \%$ uncertainty \cite{endt_cs}. This may still be the case when the data are sufficiently informative, but the results of a single experiment should be viewed more conservatively. In particular, low angular coverage in the entrance channel elastic scattering data, the absence of any elastic scattering data in the exit channel, and transfer angular distributions with just a few points all play a role in final uncertainty that can be reported for $C^2S$.

To gain a clearer picture of the role each potential plays in the final uncertainty, the calculations for the ground state were repeated for the following cases:

\begin{enumerate}
\item Uncertainty in just the entrance channel potential parameters.
\item Uncertainty in both the entrance and exit channel potential parameters.
\item Uncertainty in the entrance, exit, and bound state potential parameters.
\end{enumerate}

\begin{figure}
  \hspace*{-2cm}
  \includegraphics[width=.4\textwidth]{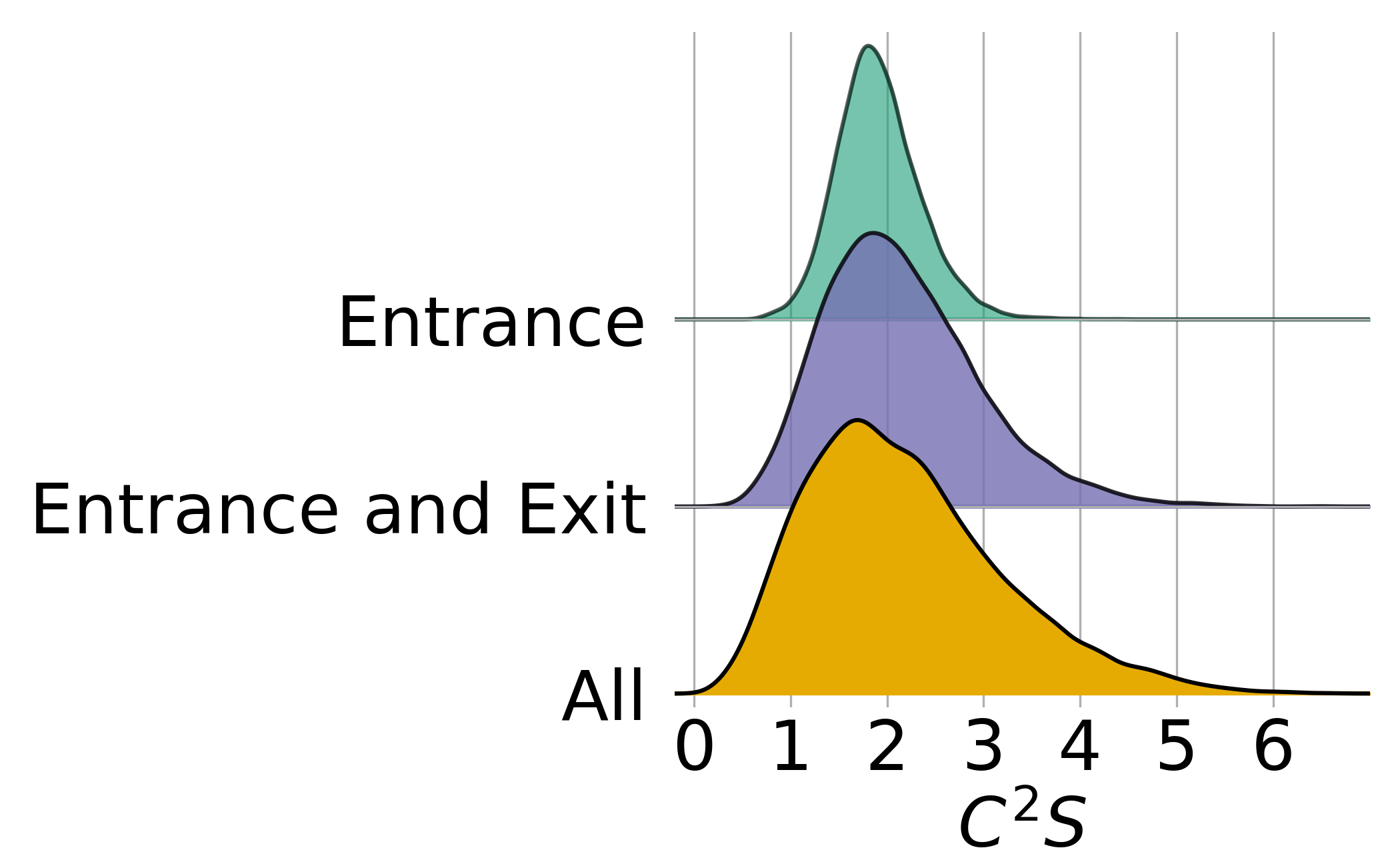}
  \caption{\label{fig:ridge} Ridge line plot that compares the KDE distributions for the ground state $C^2S$ when there is variation in the entrance potential; entrance and exit potentials; and in the entrance, exit, and bound state potentials. The percentage uncertainties go from $24 \%$, $42 \%$, and $56 \%$, respectively.}
\end{figure}

For case one we find the lowest uncertainty with $C^2S = 1.88^{+0.44}_{-0.37}$ ($\approx \! 24 \%$).
Case two is the same model used for all of the states in Sec.~\ref{sec:results}. This gives $C^2S = 2.06^{+0.87}_{-0.68}$ ($\approx \! 42 \%$). 
Case three first requires that we specify the priors for the radius and diffuseness parameters of the bound state potential. Analogously to the exit channel, which also lacks data to directly constrain these parameters, we assign $V_{\textnormal{Bound}} \sim (\mu_{\textnormal{central}, k}, \{ 0.10 \mu_{\textnormal{central}, k}\}^2)$. Again, $k$ is an index that runs over the radius and diffuseness parameters, and ``central" refers to $r = 1.25$ and $a=0.65$. This case has the largest final uncertainty with $C^2S = 2.04^{+1.15}_{-0.85}$ ($\approx \! 56 \%$).
The comparison between the final distribution for the spectroscopic factors obtained for just the entrance channel, entrance channel and exit channel, and all of the potentials including the bound state are shown in Fig. \ref{fig:ridge}. This demonstrates the strong dependence of $C^2S$ on each of these potentials.

These results point toward ways to improve the precision of $C^2S$. Examination of the correlations in the posterior samples in Fig.~\ref{fig:gs_corner} show that the imaginary radius in the exit channel is the parameter responsible for much of the uncertainty in $C^2S$. The samples for the exit channel also show little intra-potential correlation between the parameters. This is expected since the only data that could inform these parameters are in the transfer channel. If elastic data for the exit channel were available, then the proper parameter correlations could be inferred, thereby, reducing the uncertainty in the extracted spectroscopic factors. This could bring the uncertainty closer to the roughly $24 \%$ seen in the case when just the entrance potential is considered.

Bound state parameter dependence could have significant impact on astrophysical applications as well. In these applications, the extraction of $C^2S$ is an intermediate step towards calculation of quantities relevant to astrophysics such as particle partial widths and direct capture
cross sections. It was noted in Ref.~\cite{bertone} that it is essential to use the same bound state parameters for both the extraction of $C^2S$ and calculation of the direct capture cross section or partial width. This procedure was found to significantly reduce the final uncertainties on these quantities. If the bound state parameters are included in a Bayesian model to extract $C^2S$, then it becomes possible to calculate these quantities not only using the same bound state parameters, but using fully correlated, statistically meaningful samples informed directly by the transfer reaction measurement. Future work should investigate the effectiveness and impact of determining partial widths and direct capture cross sections using a Bayesian framework.        

\subsection{Nuclear Structure of $^{69} \textnormal{Cu}$}

Structure properties of $^{69} \textnormal{Cu}$ are also influenced by our results. The occupancy of orbitals tends to be higher than
expected for both the open $pf$ orbitals and for the closed $1f_{7/2}$ proton shell.
In order to propagate the uncertainties from each $C^2S$, we use the MCMC samples to
construct a KDE for each state. From these densities we pull $10^5$ samples to
estimate the occupancy:

\begin{equation}
  \label{eq:occ}
  n = \sum_{i}^N C^2S_i,
\end{equation}

where $i$ refers to each of the $N$ states considered in the sum. Similarly,
the energy of the $1f_{7/2}$ shell can be determined from:

\begin{equation}
  \label{eq:energy}
  E(1f_{7/2}) = \frac{\sum_{i}^N C^2S_i(1f_{7/2}) E_i(1f_{7/2})}{n_{1f_{7/2}}}.
\end{equation}

The occupancy above the closed shell was found to be $n_{pf} = 3.90^{+1.03}_{-1.28}$,
which is consistent but systematically higher than the value of $2.55(23)$ from the finite range calculations of the previous analysis.
For the $1f_{7/2}$ shell we have two scenarios dependent on the identity of the state at $3.70$ MeV.
If the state does not belong to the $f$ shell, we have $n_{1f_{7/2}} = 6.64^{+2.47}_{-1.79}$ and $E(1f_{7/2}) = 2.43^{+0.23}_{-0.25}$
, or if it does, $n_{1f_{7/2}} = 10.03^{+3.63}_{-2.66}$ and $E(1f_{7/2}) = 2.86^{+0.23}_{-0.26}$. 
Looking at the median value for $n_{pf}$, we would expect $n_{1f_{7/2}} = 6.10$. This may point to the $\ell=2$ assignment of the $3.70$ MeV state
being the correct one, but it must be recognized that there are still large uncertainties on all of these quantities.
Furthermore, since the optical model parameters are shared by these states, these derived values are susceptible to significant systematic shifts.
In light of this fact, these credibility intervals should be viewed as approximations. Perhaps more importantly is that if the $3.70$ state belongs to the $1f_{7/2}$, then the full strength of this shell has been observed. The shell model calculations in Ref.~\cite{pierre_paper} predict a much higher energy than $E(1f_{7/2}) = 2.86$ due to the presence of more states at higher excitation energies. A future experiment with a higher incident
beam energy that would be capable of populating these predicted higher lying states could help clarify these discrepancies.

\subsection{Comparison to Other Bayesian Studies}

It is also worthwhile to compare our methods with those of several recent publications, which have also applied Bayesian methods to optical potentials \cite{lovell_mcmc, king_dwba}.
These papers differ from our approach in a few key ways: the data come from multiple experiments, exit and entrance channels are fitted separately, transfers are calculated using finite range effects, and the prior distributions are much wider than ours ($100 \%$ of the initial global values).
{In a fully Bayesian framework, fitting the data in the entrance and exit channels separately or simultaneously is equivalent as long as the same model is used \cite{bayes}.}
While our priors are narrower, they could likely be made broader if there was elastic scattering data over a wider range of angles.
Full finite-range calculations {could be} important to include in future studies, but, {as seen in Table~\ref{tab:cs}, for this reaction the average difference is roughly $16 \%$, well within the uncertainty arising from the optical potentials.}
Including these effects will require a more efficient way to evaluate the likelihood function. Specifically, a finite-range calculation takes roughly $50$ times longer than a calculation using the zero-range approximation. For this work $2 \times 10^6$ likelihood evaluations took approximately $22$ hours, meaning the finite-range calculation would take over $1000$ hours.
As well as those differences, our results differ from those of
Ref.\cite{king_dwba} in one important aspect. Here, we confirmed the strong
correlations between optical model parameters that are expected
from historical studies \cite{hodgson1971}, and
treat them in a statistically meaningful way. We stress that our method does
not assume these correlations, but that they appear to be a consequence of the Wood-Saxon potential form factor.
On the other hand, in their comparison of frequentist and Bayesian
methods Ref.\cite{king_dwba} do not observe such
correlations, with the exception of the $V_0$ and $r_0$ anti-correlation, and ascribe their finding to the non-Gaussian
posterior distributions, which would be poorly described by the
frequentist model. We do not know the origin of this
disagreement, and suggest that it should be investigated further.

\section{Conclusions}

In this paper we have presented a method to calculate uncertainties in spectroscopic
factors and angular momentum assignments that include the uncertainty of the nuclear optical
model parameters by using Bayesian inference. We find that for the $^{70} \textnormal{Zn}(d, ^3 \textnormal{He})^{69} \textnormal{Cu}$ data analyzed here, spectroscopic factors are subject to large uncertainties that
can approach $100 \%$. If the exit channel elastic scattering is measured, this uncertainty could be reduced significantly
due to the high correlation between the exit channel imaginary radius and the spectroscopic factor.
Application of Bayesian model selection also found that there is substantial evidence in the data to
suggest an $\ell = 3$ transfer for the $3.70$ MeV state, which differs from the previous conclusion of
an $\ell = 2$ assignment. Further work is needed to resolve the ambiguity of this state. We also
find that the Bayesian approach confirms the highly correlated nature of the  
optical model potential parameters. The application of these techniques to other data sets
and global potentials could significantly improve the ability of future works to accurately
assign uncertainties to calculations involving the optical model.

\section*{Acknowledgments}

The authors would like to thank Christian Iliadis, Rafael S. De Souza, and Kiana Setoodehnia 
for their valuable input. This material is based upon work
supported by the U.S. Department of Energy, Office of Science,
Office of Nuclear Physics, under Award No. DE-SC0017799
and Contract No. DE-FG02-97ER41041.

\section*{Appendix A: Potential Ambiguities for $^{70} \textnormal{Zn}(d, d)^{70} \textnormal{Zn}$}

As discussed in Sec.~\ref{sec:amb_pots}, the continuous and discrete ambiguities pose serious issues to any uncertainty quantification of
optical model potentials. It is worthwhile to demonstrate that these issues occur naturally in our data set, and that they are not
consequences of the Bayesian methods developed throughout this paper. We do this by using artificially generated data that possess the same
characteristics of our measured data points. In particular, we use limited angular coverage, $\theta_{c.\!m.} \leq 50^{\circ}$, and equivalent
statistical uncertainties of $10 \%$. The data are generated by randomly perturbing the theoretical cross section from FRESCO according to:

\begin{equation}
  \label{eq:art_data}
  \frac{d \sigma}{d \Omega}_{\textnormal{Artificial}, i} = \frac{d \sigma}{d \Omega}_{\textnormal{Fresco}, i} + \epsilon_i, 
\end{equation}

where $i$ refers to the center of mass angle and $\epsilon_i$ is defined as

\begin{equation}
  \label{eq:epsilon}
  \epsilon_i \sim \mathcal{N} \bigg(0, \bigg(0.10 \, \frac{d \sigma}{d \Omega}_{\textnormal{Fresco}, i}\bigg)^2 \bigg).
\end{equation}

\begin{figure}
  \centering
  \includegraphics[width=.45\textwidth]{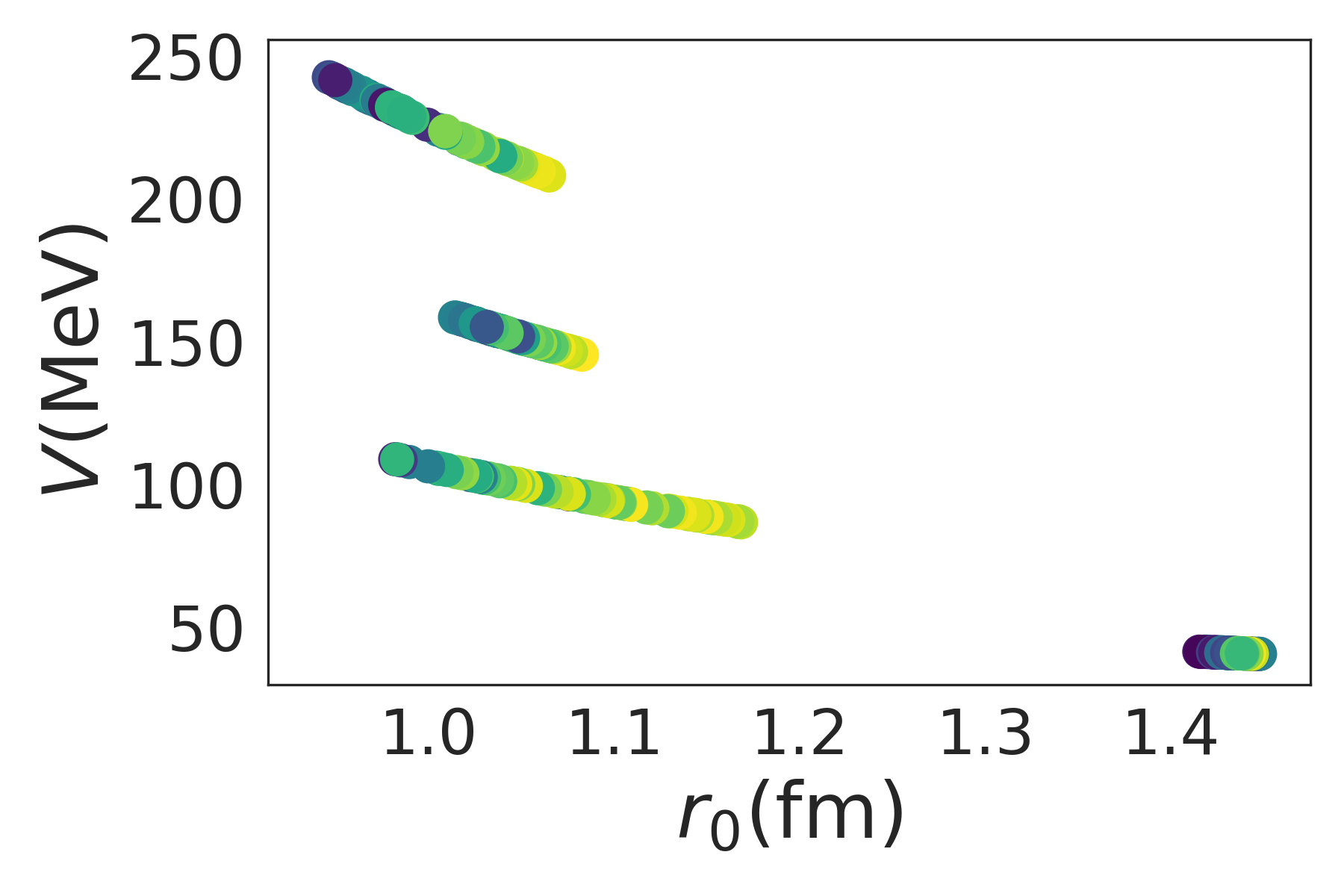}
  \caption{\label{fig:es_amb} The $V$ and $r$ values from the exploration of the maximum likelihood solutions. The colors correspond
  to the likelihood values. As the colors get brighter, the values become more likely. The continuous and discrete ambiguity can bee seen clearly.}
\end{figure}

Using a maximum likelihood estimate, 4 different values were used for the real potential, $V = 40, 80, 150, 200$ MeV. For each of these starting values
100 minimization runs were performed, with each run consisting of a minor perturbation of all of the starting parameters. This procedure effectively probes
the continuous ambiguity for each family of $V$. These results are plotted in Fig.~\ref{fig:es_amb}, which shows each solution's value of $V$ and $r_0$. Their color is based on the minimized value, with the least likely being darker and the most likely being brighter. This result shows that the two ambiguities will
be present even in an idealized situation, and that any effort to quantify uncertainties must be able to address the issues they create. 

\section*{Appendix B: $a$ parameter for Ensemble Sampling}

{
As mentioned in the main text, all of our MCMC calculations used an ensemble sampler with a stretch
move parameter set to $a=2$. We now detail the meaning of this parameter as defined in Ref.~\cite{ensemble_mcmc}.

A single step of the entire ensemble is taken by updating the positions for each walker individually. The stretch move does this for a walker
$k$ by selecting at random another walker from the ensemble and proposing an update of the form:

\begin{equation}
  \label{eq:step_prop}
  X_k(t) \rightarrow Y = X_j + z(X_k(t) - X_j),
\end{equation}

\noindent where $X_k(t)$ is the position of walker $k$ at step $t$, $X_j$ is the position of the randomly selected walker in the ensemble ($k \neq j$), and $z$ is a scaling variable drawn from the distribution
$g(z)$. This distribution is defined by the single free parameter, $a$, and is given
by the function:

\begin{equation}
  \label{eq:g_z}
  g(z) \propto
  \begin{cases}
    \frac{1}{\sqrt{z}} & \textnormal{if} \, z \in [\frac{1}{a}, a], \\
    0 & \textnormal{otherwise}.
  \end{cases}
\end{equation}

\noindent The proposed position $Y$ in a parameter space of $N$ dimensions is then accepted with probability:

\begin{equation}
  \label{eq:accept}
  q = \textnormal{min}\bigg(1, z^{N-1}\frac{p(Y)}{p(X_k(t))}\bigg).
\end{equation}

It is possible to improve the performance of the sampler by adjusting $a$. Lower values will tend to increase the
number of accepted proposals, while higher values will tend to decrease them. All calculations in this work had
acceptance fractions between $0.2$ and $0.5$, indicating the choice of $a=2$ was adequate \cite{emcee}.

}
\section*{Appendix C: Description of Nested Sampling}

We give a brief overview of the nested sampling algorithm here, but Ref.~\cite{speagle2019dynesty,skilling2006,skilling2004}
should be consulted for a more detailed explanation.
The idea of nested sampling is to estimate the evidence
by defining a prior mass, $X$, defined by an integral over the priors, $P(\theta)$, with respect to the likelihood, $\mathcal{L}$:

\begin{equation}
  \label{eq:prior_mass}
  X(\lambda) = \int_{\mathcal{L} > \lambda} P(\theta) d\theta.
\end{equation}

As $\lambda$ increases, $X$ decreases from $1$ to $0$. This definition allows the evidence integral to be written:

\begin{equation}
  \label{eq:evidence_mass}
  Z = \int_0^1 \mathcal{L}(X) dX \approx \sum_{i=1}^m \frac{1}{2}(X_{i-1} - X_{i+1}) \mathcal{L}_i,
\end{equation}

where the sum comes from the application of the trapezoid rule. The quantity $\frac{1}{2}(X_{i-1} - X_{i+1})$ is
also referred to as the weight, $w_i$.
Thus, the algorithm becomes:
\begin{enumerate}{}{}
\item Set $Z_0 = 0$ and $X_0 = 1$.
\item Draw $n$ live points from the prior distributions. 
\item Select the live point with the lowest value of $\mathcal{L}_i$, and calculate $\ln X_i \approx \frac{i + \sqrt{i}}{n}$.
\item Add the weighted sample $\mathcal{L}_iw_i$ to $Z$.
\item Update the selected live point by drawing a new point from the prior satisfying $\mathcal{L}_{i + 1} \geq \mathcal{L}_i$.
\item Repeat.
\end{enumerate}

The number of samples, $m$, required for an accurate determination of $Z$ can be
estimated from the upper bound at a step by: $\Delta Z_i \approx \mathcal{L}_{\textnormal{max}} X_i$,
where $\mathcal{L}_{\textnormal{max}}$ is the highest likelihood value of the remaining live points.

\newpage

\bibliography{paper}

\end{document}